\documentclass[prx,aps,showpacs,twocolumn, preprintnumbers,
amsmath,amssymb,superscriptaddress,longbibliography]{revtex4-2}
\usepackage[english]{babel}
\usepackage{amsmath,amssymb,amsfonts}
\usepackage{graphicx}
\usepackage[colorlinks=True,linkcolor=red,citecolor=blue,urlcolor=blue]{hyperref}
\usepackage{mathbbol}
\usepackage{bookmark}
\usepackage[dvipsnames]{xcolor} 
\usepackage{braket}
\usepackage{nicefrac}
\usepackage{bm}
\usepackage{bbm}
\usepackage{braket}
\usepackage{slashed}
\usepackage[normalem]{ulem}
\usepackage{array}
\usepackage{makecell}
\usepackage{soul}
\usepackage{cmap}
\usepackage{setspace}
\usepackage{amstext}			
\usepackage{bbold}
\usepackage{graphicx}
\usepackage{subfigure}
\usepackage{wrapfig}
\usepackage{esint}
\usepackage{hyperref}
\usepackage{multirow}
\usepackage{printlen}
\usepackage{placeins}
\usepackage{comment}
\usepackage{bm}
\usepackage[english]{babel}
\usepackage{fancyhdr}
\usepackage[left=1in, right = 1in, top = 1in, bottom = 1in]{geometry}
\usepackage{physics}
\graphicspath{{img/}}
\usepackage{media9}
\usepackage{subfiles}
\usepackage{soul}
\usepackage{orcidlink}

\hypersetup{
    colorlinks=true,
    linkcolor=blue,
    citecolor=blue,
    filecolor=blue,
    urlcolor=blue
}

\DeclareUnicodeCharacter{2212}{-}

\allowdisplaybreaks

\begin{document}
\title{Mie resonances in optical trapping: Their role in kinematics and back-action}

\author{Sharareh Sayyad\,\orcidlink{0000-0002-7725-7037}}
\affiliation{Department of Mathematics and Statistics, Washington State University, Pullman, Washington, USA}
\affiliation{Institute for Numerical and Applied Mathematics, University of G\"ottingen, G\"ottingen, Germany\looseness=-1}
\author{Gerd Leuchs\,\orcidlink{0000-0003-1967-2766}}
\affiliation{Max Planck Institute for the Science of Light, Erlangen, Germany}
\affiliation{Department Physik, Friedrich-Alexander-Universität Erlangen-Nürnberg, Erlangen, Germany}
\author{Vsevolod Salakhutdinov\,\orcidlink{0000-0003-0069-8851}}
\email[corresponding author: ]{vsevolod.salakhutdinov@mpl.mpg.de}
\affiliation{Max Planck Institute for the Science of Light, Erlangen, Germany}

\date{\today}

\begin{abstract}
The heating rate plays a crucial role in the decoherence of the harmonic motion of an optically levitated nanoparticle. The values of this rate vary depending on both the scattering photon rate and the kinetic energy acquired through individual photon recoils. While the combined roles of these factors have been extensively studied, the energy transfer per recoil has not been explicitly examined. This energy transfer is often approximated using a linear dipole model with coefficients \{1/5, 2/5, 7/5\} which applies in the Rayleigh limit. In this work, we analyze the evolution of energy transfer per photon recoil for low-absorption dielectric nanospheres with diameters ranging from 2 nm to 500 nm. Using a far-field approximation, we demonstrate that the Kerker condition, which enhances the alignment between incident and scattered wavevectors, may significantly reduce the energy transferred per recoil. Although this reduction is counterbalanced by the increasing scattering rate, for an individual scattering event, the reduction of recoil suggests an intrinsic suppression of back-action. Our results reveal a potential enhancement in the accuracy of estimations in tabletop experiments involving Mie particles of the considered sizes and provide guidance for the selection of optimal probe sizes and materials. Our interpretation of recoil reduction as a manifestation of back-action suppression indicates a potential pathway toward minimizing measurement-induced disturbances through engineered control of scattering directivity, potentially enabled by metamaterial-based designs.
\end{abstract}

\maketitle
\onecolumngrid
\section{Introduction}
Optical levitation has evolved into a versatile platform for controlling and probing microscopic motion, as extensively discussed in recent review articles~\cite{Pesce2020, Gieseler2021, Volpe2023}. Significant progress in identifying and quantifying experimental noise sources and measurement back-action limits~\cite{GonzalezBallestero2021} has further advanced these systems, enabling access to increasingly low-noise regimes and facilitating their application to high-sensitivity weak-force detection and precision sensing~\cite{Millen2020, Jin2024}. Employing levitated systems allows for exploring fundamental questions in physics, potentially including dark matter detection~\cite{Monteiro2020, Kalia2024, Kilian2024}, quantum measurement projection (collapse) models~\cite{Vinante2019}, and realization of microscopic entanglement~\cite{Poddubny2024, Zambon2024, Winkler2024}. These advances have stimulated intensive investigations on improving experimental accuracy in optical levitation experiments, expanding the boundaries of quantum control and noise reduction.

The key process limiting the experimental accuracy is the decoherence of the harmonic motion of the levitated oscillator. In high-vacuum environments, the dominant contribution to decoherence arises from the heating rate induced by photon recoil associated with scattering. This process significantly impacts the imprecision and back-action noises, collectively considered as the quantum noise level in experiments~\cite{GonzalezBallestero2023}. Due to its significance, the recoil heating rate has been extensively studied for dipolar scatterers in the Rayleigh regime~\cite{Jain2016, Tebbenjohanns2019, Abbassi2024}. Recent efforts stepped beyond this regime and determined the recoil heating rate in the intermediate and Mie regimes where levitated objects are larger. The importance of such studies relies on the link between the magnitude of recoil heating rate and the size parameter of scatterers as studied for silica~\cite{Seberson2020, Maurer2022} and silicon~\cite{Lepeshov2023}.

Using Mie particles in levitation experiments introduces new opportunities and challenges. This is because the
intrinsic characteristics of these particles give rise to electromagnetic resonances that alter the radiation patterns of scattered light. A particular example of such resonances was theoretically considered by Kerker, predicting the suppression of backscattering~\cite{Kerker1983, Liu2018}. Generally referred to as \emph{Mie resonances}~\cite{Kivshar2017}, these resonances affect the scattering characteristics, which exhibit exciting prospects for controlling levitated systems. Suggesting an auxiliary means, they can be used for stable particle confinement at a lower trapping potential by controlling attraction or repulsion forces as desired~\cite{Stilgoe2008, NietoVesperinas2010, Shilkin2022, Lepeshov2023, Mao2024}, and enhancing the detection efficiency by adjusting \emph{information radiation pattern} \cite{Maurer2022, Wang2025}.

Despite often being treated as a single parameter, the recoil heating rate arises from two distinct contributions: (1) the average kinetic energy transfer per photon scattering event, which is a kinematic process, and (2) the photon scattering rate, which is a temporal process. These two components are of inherently different physical nature and offer unique possibilities for experimental control. For instance, controlling the radiation pattern via self-interference, achieved through spherical mirrors, has been proposed to reduce back-action noise for dipolar scatterers~\cite{Weiser2025, Gajewski2025}. Hence, achieving independent control over these two contributions can play a critical role in minimizing decoherence and enhancing the performance of levitated systems.

In this work, we present a detailed numerical investigation of the kinematic component of the recoil heating rate for Mie particles that satisfy the \emph{first Kerker condition}~\cite{Nechayev2019, OlmosTrigo2020}. To gain qualitative insights, we analyze spherical particles made of silica, diamond, and silicon with diameters ranging from 2~nm to 500~nm. Trapping configurations were chosen to emulate typical experimental setups, utilizing unidirectional beams with either linear or radial polarization. Our findings reveal a considerable redistribution of the kinetic energy across translational degrees of freedom as particles approach the Kerker condition. This behavior underscores the significance of controlling scatterer radiation directivity for transferred energy redistribution and provides practical guidelines for selecting particle sizes, thereby delineating the limits of the Rayleigh approximation in the kinematic sense.

This work is structured as follows. In Sec.~\ref{sec:Motivation}, we cover the role of the kinematic component in the formation of the quantum noise, discuss its traditional consideration, and link it to the back-action effect. Section~\ref{sec:Methods} introduces a generic approach to compute kinematic components in far-field approximation applicable to the beams beyond the paraxial regime. Numerical results for several beam modes are presented in Sec.~\ref{sec:Results}, where we also provide the discussion and offer related remarks. Finally, we summarize our findings in Sec.~\ref{sec:summary}.

\section{Motivation}
\label{sec:Motivation}
The modern description of displacement measurement accuracy unavoidably faces quantum limitations. These limits are postulated by Heisenberg's uncertainty principle and set an upper bound on experimental accuracy, commonly dubbed \emph{standard quantum limit}~(SQL). In the case of the harmonic oscillator, the SQL is discussed in terms of the \emph{power spectral density} of the total quantum noise~($S_{\rm QN}(\omega)$), where $\omega$ is the consisting frequency. This quantity in its complete form reads (see, e.g.,~\cite{Clerk2010, GonzalezBallestero2023})
 \begin{equation}
     S_{\rm QN}(\omega)=S_{\rm IMP}(\omega)+S_{\rm BA}(\omega)+S_{\rm C\mbox{-}C}(\omega). \label{eq:SQN}
 \end{equation}
Here, $S_{\rm IMP}$ is linked to the \emph{imprecision noise} arising due to shot noise inherent to the measuring apparatus, e.g., uncontrolled fluctuations of the number of photons or electrons in the probing tool. The second term~($S_{\rm BA}$) accounts for the \emph{back-action noise}, which is related to the uncertainty measurement of the momentum (or field amplitude) due to uncontrollable recoil impinged on the observed object during the interaction with the probe. The third term~($S_{\rm C\mbox{-}C}$) represents the contribution originating from the cross-correlation between the previous two types of noises, where fluctuations in the probe correlate to back-action and vice versa; see details in~\cite{Clerk2010}. In the following, we limit the discussion to the uncorrelated system, where the cross-correlation term~($S_{\rm C\mbox{-}C}$) vanishes. Subsequently, for this general (uncorrelated) case, the total quantum noise becomes~\cite{Kimble2001, Clerk2010}
    \begin{align}
        S_{\rm QN}(\omega) = \frac{S_{\rm SQL}(\omega)}{2}
        \left(\frac{1}{{\cal K}(\omega)}+{\cal K}(\omega) \right).
        \label{eq:QNequationGeneral}
    \end{align}
Here, $S_{\rm SQL}(\omega)$ stands for the achievable SQL magnitude, and ${\cal K}(\omega)$ is the \emph{coupling constant} characterizing the extent to which the probe affects the observed system during the measurement process.

In the field of optical trapping, for the mechanical oscillator with mass $m$ and resonant frequency $\omega_R$, parameters ${\cal K}$ and $S_{\rm SQL}$ in Eq.~\eqref{eq:QNequationGeneral} read~\cite{GonzalezBallestero2023}
\begin{align}
    {\cal K}(\omega) &=\frac{4\Gamma_{0}}{\omega_{\rm R}}|X(\omega)|\, , \quad
    S_{\rm SQL}(\omega) = \frac{ |X(\omega)|  r_{0}^{2}}{ \pi \omega_{\rm R}}\,.
    \label{eq:Substitutions}
\end{align}
Here, $X(\omega)$ is related to the mechanical susceptibility $\chi(\omega)$ as $X(\omega)=m\,\omega^2_R\,\chi(\omega)$ and $r_0$ denotes the mean displacement of the oscillator in the ground state $r_0=\sqrt{\frac{\hbar}{2 m \omega_{\rm R}}}$. The rate $\Gamma_0$ (a.k.a. ``\emph{recoil heating rate}''~\cite{Jain2016, Maurer2022}, ``\emph{recoil rate}"~\cite{Gieseler2014}, ``\emph{phonon heating rate}''~\cite{Winstone2023}, ``\emph{rate of occupation number increase}''~\cite{Seberson2020}) is given as the inverse time required for the incident light to heat the mechanical oscillator by the energy of one phonon $\hbar\omega_{\rm R}$ due to the recoil at scattering. For the elastic scattering process, this rate is given by (see, e.g.,\cite{Seberson2020} and Sec.~4.2 in~\cite{Jain2017})
\begin{equation}
\Gamma_{0,\rm i} =\frac{1}{\Delta t}=\frac{\kappa_{\rm i}\,\Delta E_{\rm p}\,\nu^{\rm sc}_{_{\rm [ph/s]}}}{\hbar \omega_{\rm R}}\,,
\label{eq:Gamma0}
\end{equation}
where $\nu^{\rm sc}_{_{\rm [ph/s]}} = I \sigma_{\rm sc} / (\hbar \omega_{0})$ sets the rate of scattered photons (the number of scattering events per second), $I$ denotes the focal intensity of light, $\sigma_{\rm sc}$ is the scattering cross-section and $\omega_{0}$ is the light frequency. Here, $\Delta E_{\rm p}$ is the kinetic energy of the oscillator equal to the energy of the incident photon with the wave number $k$, i.e., $\Delta E_{\rm p}=\frac{\hbar^2 k^2}{2\,m}$. The coefficient $\kappa_{\rm i}$ accounts for the average fraction of the kinetic energy distributed over three translational degrees of motion, i.e., ${\rm i}\in \{x,y,z\}$, after the single scattering event. In the other formulation $\kappa$ is the dispersion fraction of mechanical momentum difference between the incident $k\hat{k}_{\rm in}$ and scattered $k\hat{k}_{\rm sc}$ photons' momenta with wavenumber $k=\omega_{0}/c$ (see, e.g.,~\cite{Itano1982} and Sec.~4.2 in~\cite{Jain2017}), where $\hat{k}_{\rm i}$ denotes a unit vector along $\vec{k}_{\rm i}$.
Since the heating rate $\Gamma_{\rm 0}$ characterizes the decoherence introduced by trapping light to the measurement, its magnitude draws substantial interest (see, e.g.,~\cite{Seberson2020, Maurer2022}). The studies treat $\Gamma_{\rm 0}$ as a solid parameter with no explicit distinction between the consisting components in Eq.~\eqref{eq:Gamma0}. In the current work, we focus on the quantity characterizing the kinetic energy variation of the oscillator interacting with the incident photon during an individual scattering event. In this sense, $\kappa$ is treated as a kinematic recoil indicator per scattering event rather than as a complete description of back-action noise within the full quantum measurement framework. Accordingly, we primarily analyze the magnitude of $\kappa$ as a parameter quantifying the recoil component of the scattering process.

In general, the coefficient $\kappa$ appears as a result of the calculation of the autocorrelation function of the force impinged onto the trapped harmonic oscillator (see, e.g., supplementary to~\cite{Jain2016}, Sec.~4.2 in~\cite{Jain2017}, and~\cite{Gajewski2025}), which is needed for deriving the back-action or imprecision noises. For the plane incident wave propagating along the direction $z$-axis and polarized along the $x$-direction, i.e., $\vec{k}_{\rm in}=k\,(0,0,1)^{\rm T}$, the parameter $\kappa$ is written as (see, e.g.,~\cite{GonzalezBallestero2019, Seberson2020}):
\begin{equation}
    \kappa_{\rm x,y,z}=\{\langle \hat{k}^{2}_{\rm sc,x}\rangle_{_{\rm \Omega}},\langle \hat{k}^{2}_{\rm sc,y}\rangle_{_{\rm \Omega}},\langle\hat{k}^{2}_{\rm sc, z}\rangle_{_{\rm \Omega}}+1\}\,,
\end{equation}
where $\hat{k}_{\rm sc,i}$ is the $i$th component of a unit vector along the wavevector $\vec{k}_{\rm sc}$ of the scattered photon propagation and the angle brackets $\langle...\rangle_{_{\rm \Omega}}$ stands for the averaging over full solid angle $\Omega$. The extra unity in the z-component appears due to the \emph{radiation pressure} caused by the momentum of the incident photon $\hbar\vec{k}_{\rm in}$ permanently applied to the oscillator. Assuming the scatterer to be a linear dipole with differential power $\frac{\dd P_{_{\rm dip,\,sc}}(\theta)}{ \dd \Omega} \propto 1-\sin^2\theta\cos^2\phi$ ($\theta$ is the polar angle counted from the negative direction of z, and $\phi$ is the azimuthal angle in the plane perpendicular to z, see Fig.~S1 in the Supplemental Materials~(SM)~\cite{SuppMat}), where averaging over the full solid angle results in $\{1/5,2/5,7/5\}$ for $\kappa_{\rm  x,y,z}$; see, e.g.,~\cite{Jain2016, Tebbenjohanns2019,  Seberson2020}. Attempts to extend these results to the case of the trapping Gaussian beam focused with high numerical aperture~($\rm NA$) leads to rewriting $\kappa_{\rm}$ as~\cite{Tebbenjohanns2019}
\begin{equation}
    \kappa_{\rm x,y,z}=\{\langle \hat{k}^{2}_{\rm sc,x}\rangle_{_{\rm \Omega}},\langle \hat{k}^{2}_{\rm sc,y}\rangle_{_{\rm \Omega}},\langle\hat{k}^{2}_{\rm sc, z}\rangle_{_{\rm \Omega}}+\xi^2_{_{\rm Gouy}}\}\,,
    \label{kappaWITHGouy}
\end{equation}
where $\xi_{_{\rm Gouy}}$ is an effective parameter characterizing the total phase of the incident beam $\varphi(z)=k z + \varphi_{_{\rm Gouy}}(z)$ along the propagation direction z-axis, where $\varphi_{_{\rm Gouy}}(z)$ is the Gouy phase, at the position $z$ of a trapped particle; see~\cite{Tebbenjohanns2019}, Sec.~3.3 in~\cite{GajewskiTH2024}. In experiments tracking the displacement of the scatterer's position, the phase $\varphi_{_{\rm Gouy}}(z)$ plays a central role, as it is imprinted in the scattered light, making position retrieval possible~\cite{Gittes1998,Pralle1999,Tebbenjohanns2019}. The factor $\xi_{_{\rm Gouy}}$ is included in the relation $\langle F_{\rm in,z}\rangle_{\rm t} = \xi_{_{\rm Gouy}} P_{\rm sc}/c$, between scattered power $P_{\rm sc}$, and time-averaged \emph{radiation pressure force} $F_{\rm in}$ exerted along the beam's propagation direction, and the constant $c$ is the speed of light ~\cite{Tebbenjohanns2019}. On the other hand, the radiation pressure force can be canonically written as $\langle F_{\rm in,z}\rangle_{\rm t} = ((\vec{\nabla} \varphi)_{\rm z} / k)  P_{\rm sc}/c$ (see, e.g., Sec.~14.4 in~\cite{NovotnyHecht2012} and Sec.~16.10 in~\cite{Zangwill2013}), that offers the physical interpretation of $\xi_{_{\rm Gouy}}$. The parameter $\xi_{_{\rm Gouy}}$ is an effective propagation constant which represents the effective longitudinal component of the wavevector characterizing the incident Gaussian beam; see also~\cite{Feng2001} for details.

For our further analysis, it is important to note that $\kappa_{\rm in,z}$ from Eq.~\eqref{kappaWITHGouy} can be expressed solely in terms of the angular distributions of the incident and scattered fields. Such, the photon recoil force $F_{\rm sc,z}$ for an active emitter (scattering without absorption)  can be described via the scattering angle $\theta_{\rm sc}$, taken as an azimuthal angle to the z-axis. This dependence can be formulated in terms of $\langle\cos \theta_{\rm sc}\rangle_{_{\rm \Omega}}$, which is commonly referred to as the \emph{asymmetry factor} (see, e.g.,~\cite{Rohrbach2001, Rohrbach2004, Rohrbach2005} and Sec.~3.11~\cite{Kerker1969}); see SM~\cite{SuppMat} for details. On the other hand, for a passive scatterer (absorption without scattering) the parameter $\xi_{_{\rm Gouy}}$ can be defined as the mean longitudinal projection of the wavevector $\langle k_{\rm in,z}\rangle$, i.e. $\langle \cos \theta_{\rm in}\rangle_{_{\rm \Omega}}$, averaged over the angular spectra function characterizing incident field (see~\cite{Rohrbach2001, Rohrbach2004, Rohrbach2005}). These relationships suggest that the parameter $\kappa_{\rm z}$ can be expressed in terms of the asymmetry factors, offering an alternative calculation method compared to the approach based on the Gouy phase approximation. Specifically, it can be defined as $\langle (\cos \theta_{\rm sc} - \cos \theta_{\rm in})^{2}\rangle_{_{\rm \Omega}}$, which yields $\kappa_{\rm in,z}$ from Eq.~\eqref{kappaWITHGouy}, since $\langle \cos \theta_{\rm in}\rangle_{_{\rm \Omega}}=0$ when averaged dipole scattering function $\frac{\dd P_{_{\rm dip,,sc}}(\theta)}{ \dd \Omega}$.

Obtaining $\kappa$ in a plane wave approximation enables us to account for the impact of the Gaussian beam along the propagation direction. However, the approach is limited to the case of the incident plane wave with the parameter $\xi_{_{\rm Gouy}}$  relevant to the beam propagation direction z. In fact, modeling the incident field as a plane wave inherently neglects perturbations associated with the lateral momentum components $k_{\rm in,\,x}$ and $k_{\rm in,\,y}$, which is only a valid approximation for the field distribution at the exact coordinate of the focus. This restriction causes $\kappa_{\rm x}$ and $\kappa_{\rm y}$ to remain unperturbed. Thus, if one has an interest in the general evolution of the parameters $\kappa_{\rm i}$, and the components of the recoil heating $\Gamma_{\rm 0, i}$, the general treatment beyond the incident plane wave approximation should be considered.

Given the central role of $\kappa$ in defining $\Gamma_0$ and $S_{\rm BA}$, $\kappa$ directly influences the decoherence introduced by the trapping light. Modifying parameter $\kappa$ can significantly influence back-action noise, thereby affecting measurement precision. Representing energy redistribution following an individual scattering event, $\kappa$ characterizes the kinematic process. In this way, a better understanding of $\kappa$ can shed light on the mechanism of reducing the undesired heating due to photon recoil and the corresponding reduction of the back-action effect. This viewpoint stimulates our interest in the evolution of $\kappa$ depending on different setup conditions and various scatterers. In this regard, we have generalized the calculation presented in Sec.~4.2 in~\cite{Jain2017}, which was obtained for the autocorrelation function of the instantaneous scattering force. The approach involves incorporating both the incident and scattered photon momenta, which results in the expression for back-action noise as $S_{\rm BA} \propto \langle (\hat{k}_{\rm in, i}-\hat{k}_{\rm sc, i})^{2} \rangle_{_{\rm \Omega}}$, see the comment \footnote{An alternative approach for calculating $S_{\rm BA}$ was recently proposed, see~\cite{Gajewski2025, Abbassi2024}. It suggests calculating the force autocorrelation function in the frequency domain and utilizes Green's functions and Fisher information representations.}. The given averaging $\langle ... \rangle_{_{\rm \Omega}}$ highlights the importance of the scattered radiation distribution as a weighted function, which directs our attention to its modification due to possible Mie resonances. In the subsequent section, we present the details of the calculation approach we followed.

\section{Methods}
\label{sec:Methods}
\begin{figure}
    \centering \includegraphics[width=0.7\linewidth]{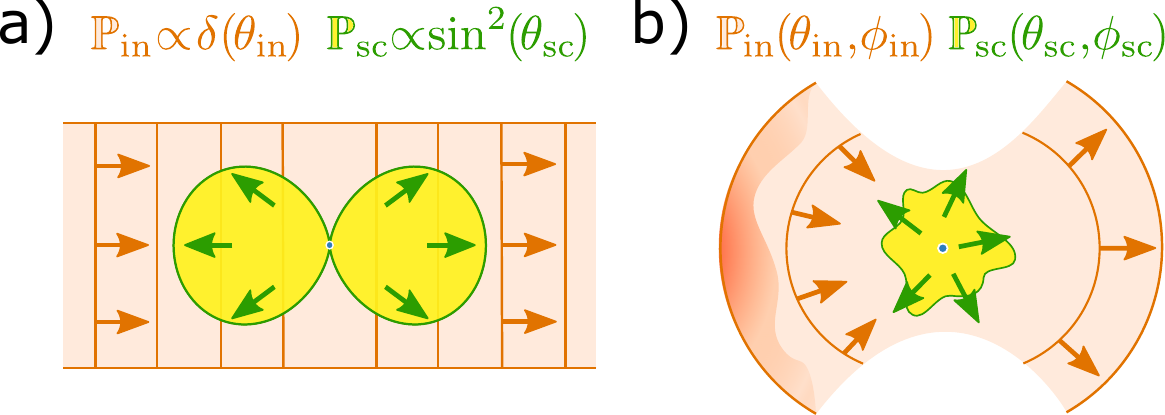}
    \caption{\textbf{Comparison between the conventional and proposed approach for treating arbitrary radiation patterns.} Schematic illustration of the conventional consideration with the incident plane wave with $\mathbb{P}_{\rm in}\propto \delta(\theta_{\rm in})$ and dipolar scatterer $\mathbb{P}_{\rm sc}\propto \sin(\theta_{\rm sc})$ at a) and generic case of arbitrary functions $\mathbb{P}_{\rm in}$ and $\mathbb{P}_{\rm sc}$ at b). All the waves, except the plane incident wave at a), have a spherical wavefront.}
    \label{fig:Concept}
\end{figure}

The calculation approach we follow in our work is based on the seminal work~\cite{Itano1982}, which discusses the averaged kinetic energy acquired by the elementary scatterer (linear or circular dipole) under the influence of subsequent inelastic light scattering. Authors consider an incident plane wave with a wave vector $\vec{k}_{\rm in}$ and scattered light with wavevector $\vec{k}_{\rm sc}$ and differential power $\dd P_{\mathrm{sc}}(\theta,\phi)/\dd\Omega$ (hereafter referred to as $P^{^{_\Omega}}_{\mathrm{sc}}(\theta,\phi)$) given in the far-field, see illustration in Fig.~\ref{fig:Concept}. The resultant parameters $\kappa_{\rm i}$ was defined as $\kappa_{\rm i}=\langle (\hat{k}_{\rm in, i}-\hat{k}_{\rm sc, i})^{2}\rangle_{_{\rm \Omega}}$, that can be written as (see~\cite{Itano1982,Seberson2020} and Sec.~4.2 in~\cite{Jain2017})
\begin{equation}
\kappa_{\rm i}=\int_{\Omega}\,\mathbb{P}_{\rm sc}(\theta,\phi) \,(\hat{k}_{\mathrm{in}}-\hat{k}_{\mathrm{sc}})_{\rm i}^{2}\,\dd \Omega\,,
\label{eq:oldDeltaK2}
\end{equation}
where $\mathbb{P}_{\rm sc}(\theta,\phi)$ is the probability density function corresponding to the normalized differential power distribution in the far field, and which integration is carried out over the full solid angle $\dd \Omega =\sin\theta\, \dd \theta \dd \phi$\,. 
Nevertheless, this commonly accepted formulation is a particular case in the sense that the incident beam is set to be a plane wave. Within this formulation, the wave vector $\vec{k}_{\rm in}$ propagation direction is given by angles ($\theta_{\rm in}, \phi_{\rm in}$) (``${\rm in}$'' stands for the ``incident'') and the direction of the incident beam is determined by $\delta(\theta_{\rm in}, \phi_{\rm in})$.
Subsequently, the incident field polarized along $\vec{e}$ direction can be defined in the k-space as $\vec{E}_{\rm in}=\vec{e}\,E_0\,\delta (\vec{k}\,(\theta_{\rm in},\phi_{\rm in}))$ (see, e.g., Sec.~III.C in~\cite{Rohrbach2002}). This subtle aspect has not been frequently emphasized, as the propagation direction of the incident plane wave is typically chosen as the reference axis for the sake of simplicity. However, a more general formulation of the problem requires characterization of incident light beyond the assumption of unidirectional propagation, see, e.g., \cite{Anishur2021, Sinha2022}; thereby implying the necessity of employing distribution functions characterizing the incident light.

At the next stage, we follow a generalized formulation, introducing the joint probability of two independent events characterizing incident photon income and scattered photon outcome. Each one is described by respective probability density function, $\mathbb{P}_{\rm in}$ and $\mathbb{P}_{\rm sc}$ (see Fig.~\ref{fig:Concept}(b)), given by the normalized differential power as $\mathbb{P}_{\rm i}(\theta_{\rm i},\phi_{\rm i}) = P^{^{_\Omega}}_{\rm i}(\theta_{\rm i},\phi_{\rm i})/\int_{\Omega_{\rm i}}P^{^{_\Omega}}_{\rm i}(\theta_{\rm i},\phi_{\rm i}) \dd \Omega_{\rm i}$, with ${\rm i}\in(``{\rm in}",``{\rm sc}")$, see \cite{PoyntingVector}.

As a result, the general form for $\kappa_{\rm i}$ becomes 
\begin{align}
    \kappa_{\rm i}=
    \int_{\Omega_{\mathrm{in}}} \int_{\Omega_{\mathrm{sc}}}\,&
    \mathbb{P}_{\rm in}(\theta_{\rm in},\phi_{\rm in})\, \mathbb{P}_{\rm sc}(\theta_{\rm sc},\phi_{\rm sc})\,
    \left(\hat{k}_{\mathrm{\rm in}}-\frac{\sigma_{\rm sc}}{\sigma_{\rm ext}}\hat{k}_{\rm sc}\right)_{\rm i}^{2} \,\dd{\Omega_{\rm sc}}\, \dd{\Omega_{\rm in}}\,,
    \label{eq:DeltaK2}
\end{align}
where $\sigma_{\rm sc}$ and $\sigma_{\rm ext}$ correspond to the scattering and extinction cross-sections, respectively. Further, we work in the frame of the absorptionless nanoparticles, claiming that we are in a regime where $\frac{\sigma_{\rm sc}}{\sigma_{\rm ext}}\approx 1$. In this limit, negligible absorption maximizes the probability of scattered-photon events. A nonzero imaginary part of the refractive index of the scatterer introduces absorption, which reduces the fraction of photons re-emitted into the scattering channel and thus proportionally decreases the mean scattered momentum via the factor $\frac{\sigma_{\rm sc}}{\sigma_{\rm ext}}$, in addition to the heating effects discussed below in Sec.~\ref{sec:Results}. For the parameters considered here, this ratio remains close to unity, so the scattering process dominates. The elements of the solid angles $\dd{\Omega_{\rm in}}$ and $\dd{\Omega_{\rm sc}}$ in Eq.~\eqref{eq:DeltaK2} are given in two independent spherical coordinate systems sharing the same basis vectors. Note that in Eq.~\eqref{eq:DeltaK2}, writing the total probability $\mathbb{P}_{\rm tot}$, as
 \begin{equation}
 \mathbb{P}_{\rm tot} (\theta_{\rm in},\phi_{\rm in}, \theta_{\rm sc},\phi_{\rm sc}) = \mathbb{P}_{\rm in}(\theta_{\rm in},\phi_{\rm in})\, \mathbb{P}_{\rm sc}(\theta_{\rm sc},\phi_{\rm sc}) \, ,
 \label{eq:ptot}
 \end{equation} 
where this factorization represents a product form of the joint angular distribution of incident and scattered momentum directions in the far-field. Within this representation, angular correlations associated with the coherent field structure are not retained, so the incident and scattered angular variables are treated as effectively uncorrelated. Residual deviations arising from the omission of near-field contributions are quantified below and discussed in detail in the SM~\cite{SuppMat} and do not affect the conclusions of this work. Assuming $\mathbb{P}_{\rm in}\propto \delta(\theta_{\rm in}) \delta(\phi_{\rm in})$ in Eq.~\eqref{eq:DeltaK2}, we reproduce the canonical convention in Eq.~\eqref{eq:oldDeltaK2} for the case of the incident plane wave, see, e.g.,~\cite{Seberson2020} and the supplementary to~\cite{Jain2016}.

\begin{figure}
    \centering
    \includegraphics[width=0.7\linewidth]{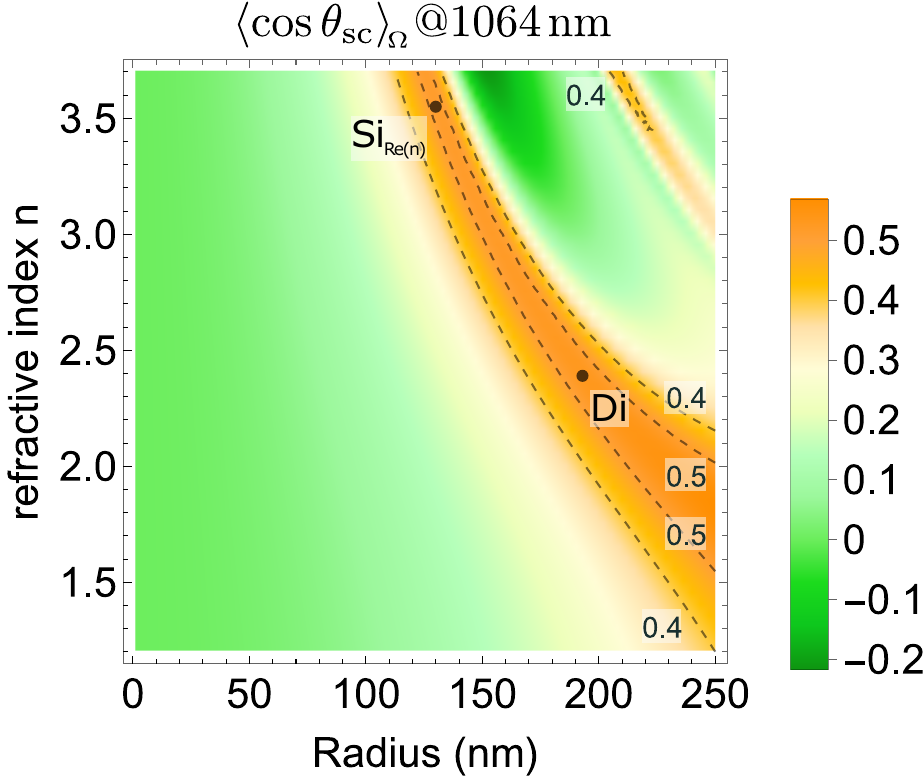}
    \caption{\textbf{Magnitude of the asymmetry parameter $\langle \cos\theta_{\rm sc} \rangle_{_{\rm \Omega}}$ depending on refractive index $n$ and radius of the dielectric nanosphere.} The refractive index is set to be purely real-valued. The distribution is calculated for the plane incident wave with $\lambda=1064~{\rm nm}$ according to Eq.~\eqref{eq:CosMean} given in the SM~\cite{SuppMat}. Black points represent the position of the highest directivity obtained for diamond (Di) and the real part of silicon (Si).}
    \label{fig:Cos}
\end{figure}

A fundamental characteristic of the scattering function $\mathbb{P}_{\rm sc}(\theta_{\rm sc},\phi_{\rm sc})$ is its explicit dependence on the magnitude and angular distribution of the incident electric field $\vec{E}_{\rm in}(\theta_{\rm in}, \phi_{\rm in})$. This inherent dependence results in a correlation between the scattered and incident fields. To explicitly incorporate this correlation, we utilize a decomposition of the incident field into a set of plane wave components, each associated with a known scattered-field solution $\vec{E}_{\rm sc}(\theta_{\rm sc},\phi_{\rm sc},\theta_{\rm in}, \phi_{\rm in})$, determined analytically via Mie theory (see SM~\cite{SuppMat}). Consequently, the resulting scattering differential power $P_{\mathrm{sc}}^{^{_\Omega}}(\theta_{\rm in}, \phi_{\rm in})$ emerges as a coherent superposition of these individual scattered fields, yielding the expression $P_{\mathrm{sc}}^{^{_\Omega}}(\theta_{\rm sc}, \phi_{\rm sc})\propto |\int \vec{E}_{\rm sc}(\theta_{\rm sc},\phi_{\rm sc},\theta_{\rm in}, \phi_{\rm in}) \dd{\Omega_{\rm sc}} |^{2}$. This formulation provides an explicit definition of the scattering function for an arbitrary incident field, making it directly applicable in subsequent computations involving Eq.~\eqref{eq:DeltaK2}.

In this context, determining $\mathbb{P}_{\rm in}$ from the angular spectrum of $\vec{E}_{\rm in}(\theta_{\rm in}, \phi_{\rm in})$ in the far-field omits the near-field contribution in the vicinity of the focal region where the scatterer is located. Consequently, the model is of an approximate nature. This omission introduces a discrepancy in the determination of the parameter $\langle \hat{k}^{2}_{\rm in,z}\rangle$. In the particular case of a pure dipole emitter illuminated by a focused Gaussian beam, the relative error with respect to the conventional definition of $\langle k ^{2}_{\rm in,z}\rangle$ (also denoted as $\xi^2_{_{\rm Gouy}}$ in Eq.~\eqref{kappaWITHGouy}) increases from 0\% at ${\rm NA} = 0$ to 13\% at ${\rm NA} = 0.9$ (see SM~\cite{SuppMat} for details). Since the near-field contribution decays rapidly with distance from the focus, we regard these values as representing the maximum error tolerance of our model for estimating the incident-field contribution. These estimates correspond to the Gaussian beam \emph{filling factor} used in this work, discussed in detail below, and tend to be overestimated for both smaller and larger filling factors.

Analysis of the integral in Eq.~\eqref{eq:DeltaK2} show that minimizing the $\kappa_{\rm i}$ requires the $\vec{k}_{\rm in}$ and $\vec{k}_{\rm sc}$ to be co-directed. Since the incident beam is directed predominantly along the propagation direction z, the scattered light has to be directed along the positive z axis. This directivity magnitude can be characterized with the asymmetry factor $\langle\cos \theta_{\rm sc}\rangle_{_{\rm \Omega}}$. Being evaluated for the nanospheres of absorptionless materials with the refractive index $n\in[1.2,3.7]$, the asymmetry factor is presented in Fig.~\ref{fig:Cos}. The greater magnitude with  $\langle\cos \theta_{\rm sc}\rangle_{_{\rm \Omega}}>0.5$ can be achieved when the Kerker condition of the first type is met, see SM~\cite{SuppMat}. 

In the evaluation of Fig.~\ref{fig:Cos}, we assume that the radius of the sphere varies in the range $\rm{R}\in[1~{\rm nm},250~{\rm nm}]$ and the wavelength of the incident light is $\lambda=1064~{\rm nm}$. 
These parameters are used throughout the following work. 
The choice of the nanoparticle size range is dictated by the diffraction limit, specifically the condition \( R \lesssim \lambda / 2 \), which provides a reference scale separating the Rayleigh scattering regime from the geometrical optics regime, while the particle sizes considered here fall within the intermediate Mie regime. This choice also reflects an optimal compromise between computational effort and physical insight. A similar choice was adopted in \cite{Seberson2020}, which serves as a related reference for our study.

In order to implement the formulation given in Eq.~\eqref{eq:DeltaK2}, we assume two different distributions of the incident electric field determined in the far field, which defines $\mathbb{P}_{\rm in}(\theta_{\rm in},\phi_{\rm in})$ (see SM~\cite{SuppMat}). The first one is a Gaussian beam with filling factor $f_0=0.8$, and the second one is a superposition of Hermit-Gaussian modes ${\rm HG}_{\rm 01}$ and ${\rm HG}_{\rm 10}$ resulting in the radially polarized beam with $f_0=0.7$. The choice of the filling factors magnitudes allows at least 95\% of the total incident intensity to be collected by the focusing objective of the specified numerical aperture. The NA was chosen to be varied in the range of ${\rm NA}\in[0.4,0.9]$ and defines the polar angle in the span covered by $\Omega_{\rm in}$ with $\phi_{\rm in}\in[0,2\pi]$ and $\theta_{\rm in}\in[0,\arcsin({\rm NA})]$. 

As noted previously, to calculate the function $\mathbb{P}_{\rm sc}$, we used an extended framework of the Mie theory obtained for linearly polarized incident plane wave, see SM~\cite{SuppMat}. The resultant scattering field depends on the radius of the dielectric scatterer, its refractive index, the vectorial electric field of the incident light, and the NA of the focusing objective.
Defining all these parameters and substituting them in Eq.~\eqref{eq:DeltaK2} allows us to obtain $\kappa$ for the three translational motion degrees. Due to excessive computational expenses, the calculations were carried out on a grid with 154 points equidistantly distributed on the plane NA\,$\cross$\,R (11\,$\cross$\,14). In the following section, we summarize results obtained for $\kappa$ under the prescribed setup conditions and for scatterers made of three different materials.

\section{Results and Discussion}
\label{sec:Results}
\begin{figure*}
    \centering
    \includegraphics[width=0.98\linewidth]{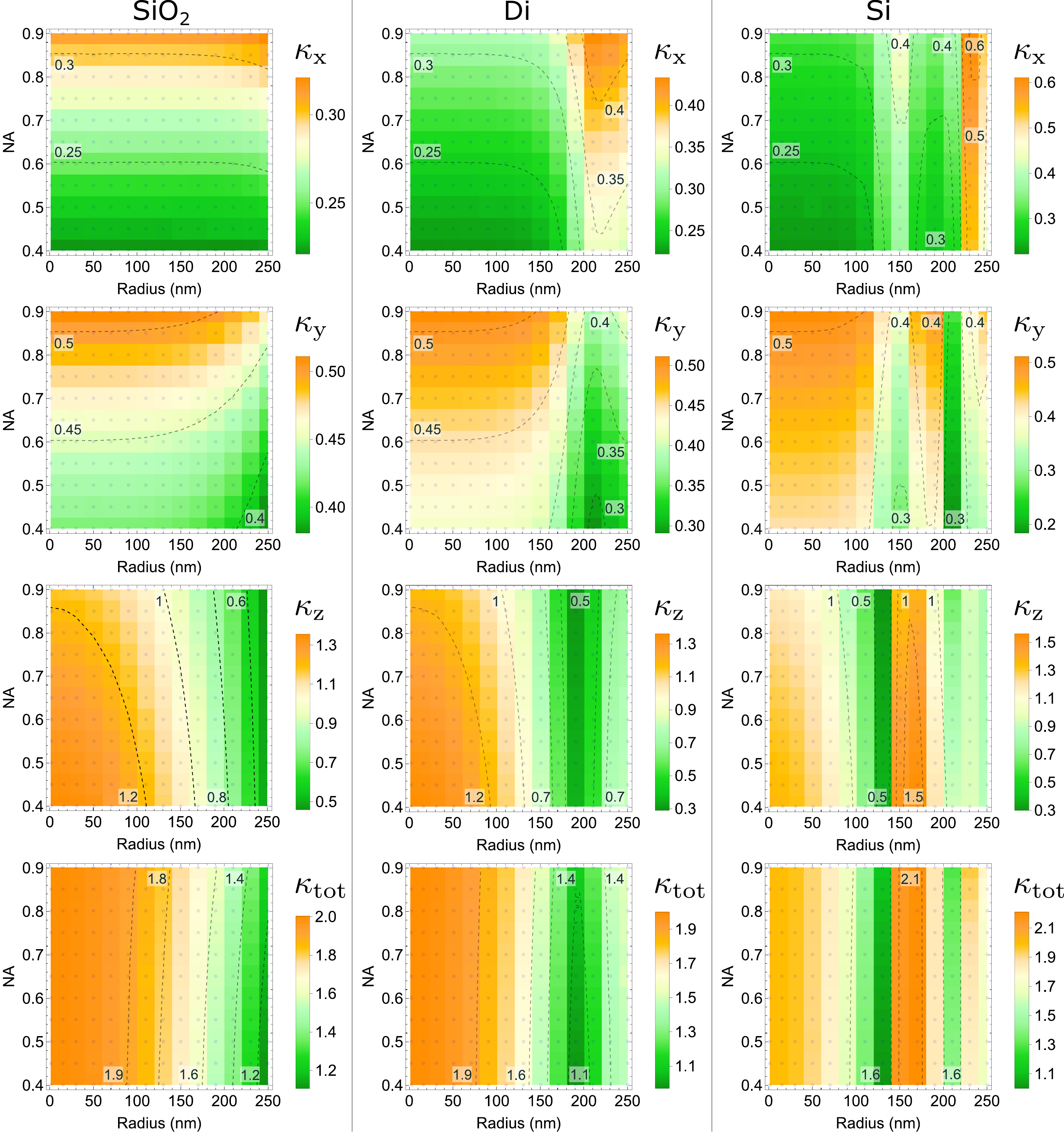}
    \caption{\textbf{The magnitude of $\kappa_{\rm i}$ for the linearly polarized incident Gaussian beam with $f_{0}=0.8$.} Columns, from left to right, correspond to the spherical particles made of silica (${\rm SiO}_{2}$), diamond (Di), and silicon (Si). The rows, from top to bottom, represent x,~y,~z components and their sum (tot), respectively. The contour lines are obtained by linear spline interpolation of the values calculated with Eq.~\eqref{eq:DeltaK2} at the coordinates marked with dark points.}
    \label{fig:Sec7kappaLinear}
\end{figure*}

\begin{figure*}
    \centering
    \includegraphics[width=0.98\linewidth]{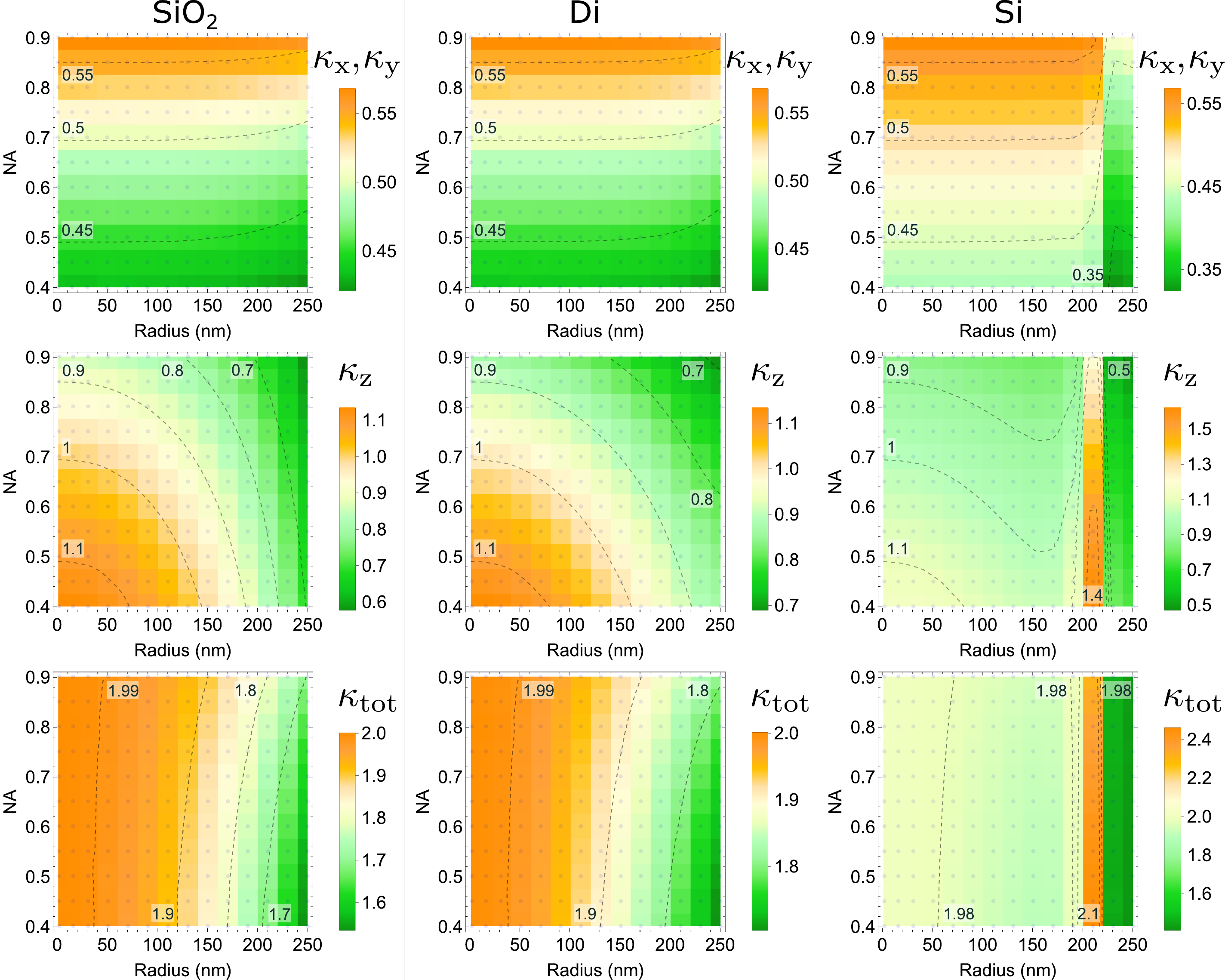}
    \caption{\textbf{The magnitude of $\kappa_{\rm i}$ for the radially polarized incident beam with $f_{0}=0.7$.} Columns, from left to right, correspond to the spherical particles made of silica (${\rm SiO}_{2}$), diamond (Di), and silicon (Si). The rows, from top to bottom, represent x (equally y),~z components and their sum (tot), respectively. The contour lines are obtained by linear spline interpolation of the values calculated with Eq.~\eqref{eq:DeltaK2} at the coordinates marked with dark points.}
    \label{fig:Sec7kappaRadial}
\end{figure*}

Having discussed our approach to calculate $\kappa_{\rm i}$ using various radiation probabilities, we here present our numerical results for $\kappa$. Our results consist of various distributions of $\kappa_{\rm i}$ and their sum -- $\kappa_{\rm tot}=\sum_{\rm i} \kappa_{\rm i}$ with $ \rm{i}\in \{x,y,z\}$, for three different cases: 1) silica ``${\rm SiO_{2}}$'', 2) diamond ``Di'', and 3) silicon ``Si'' nanosphere, illuminated by linear Gaussian or radially polarized beams with wavelength of 1064~{nm}. For this wavelength, the corresponding refractive indices were chosen to be $n_{\rm SiO_2@1064nm} = 1.45$, $n_{\rm Di@1064nm} = 2.39$, and $n_{\rm Si@1064nm} = 3.5548 + 82 \cdot 10^{-6}\,i$. The distributions of $\kappa$ for these three refractive indices are presented in Fig.~\ref{fig:Sec7kappaLinear} and Fig.~\ref{fig:Sec7kappaRadial} in the left, middle, and right columns, respectively. Given the close connection between $\kappa_{\rm z}$ and the magnitude of $\langle\cos \theta_{\rm sc}\rangle_{{\rm \Omega}}$ described earlier, there is a clear correlation between the $\kappa_{\rm z}$ minima in Fig.~\ref{fig:Sec7kappaLinear} (distributions in Fig.~\ref{fig:Sec7kappaRadial} are discussed separately) and the maxima of $\langle\cos\theta_{\rm sc}\rangle_{\Omega}$ in Fig.~\ref{fig:Cos}. This correlation provides a basis for elucidating a general trend in $\kappa$ distributions from the behaviour of $\langle\cos\theta_{\rm sc}\rangle_{\Omega}$. Since the selected refractive index values are nearly equidistant along the axis shown as the ordinate in Fig.~\ref{fig:Cos}, analysing variations in the $\kappa_{\rm z}$ distributions for these materials allows us to infer and extrapolate the qualitative behaviour to other materials with intermediate refractive indices also presented in Fig.~\ref{fig:Cos} and possessing a low imaginary part, ${\rm Im}[n] \ll 1$. Therefore, the subsequent discussion may also appeal to readers whose research lies beyond the specific cases considered in this work.

\paragraph{Incident linearly/circularly polarized light.} 
Figure~\ref{fig:Sec7kappaLinear} displays distributions of $\kappa_{\rm i}$'s and $\kappa_{\rm tot}$ as a function of radius and NA for silica nanosphere illuminated with linearly polarized incident light. Due to the present axial symmetry in the x-y plane for the cases with incident circular polarization, the x or y components of $\kappa$ can be evaluated as an averaged sum of $\kappa_{\rm x}$ and $\kappa_{\rm y}$ obtained for the incident linear polarization, i.e., $(\kappa_{\rm x}+\kappa_{\rm y})/2$. Hence, we merely focus on exploring our results for incident linear polarization.

The analysis of the distributions corresponding to silica in Fig.~\ref{fig:Sec7kappaLinear} reveals that $\kappa_{\rm tot}$ and $\kappa_{\rm z}$ gradually decrease with increasing radius, exhibiting no anomalous behavior. The lateral components $\kappa_{\rm x}$ and $\kappa_{\rm y}$ remain predominantly constant and show negligible variation within the range $R\in [1~{\rm nm}, 170~{\rm nm}]$, consistent with the Rayleigh approximation $R\rightarrow0$.

These observations can be attributed to the fact that a silica nanoparticle with a radius up to 250~{nm} does not exhibit significant changes in $\langle \cos \theta \rangle_{_{\rm \Omega}}$, as shown in Fig.~\ref{fig:Cos}, and the redistribution of radiation power with increasing radius (see Fig.~S2~(a),(b) in the SM~\cite{SuppMat}) occurs gradually. Consequently, one can employ the parameters $\kappa_{\rm i}$ obtained for a radius of $1~\mathrm{nm}$ as a reliable approximation for a wide range of radius values.

In contrast to silica, the distributions of $\kappa_{\rm i}$ obtained for diamond and silicon display noticeable changes with radius growth (see Fig.~\ref{fig:Sec7kappaLinear}~(middle and right columns)). Analyzing the total parameter $\kappa_{\rm tot}$, we observe its drastic drop near the radii corresponding to the maximally forward-directed light (about 130~{nm} for Si and 193~{nm} for diamond). For the z component, these observations are also evidently expressed, as the longitudinal (along z direction) radiation pattern is affected more with the growth of radius (see SM~\cite{SuppMat}). Besides, Fig.~\ref{fig:Sec7kappaLinear} exhibits that when approaching the size of the minima dip for $\kappa_{\rm z}$, $\kappa_{\rm x}$ and $\kappa_{\rm y}$ continue their change, and extremum of these changes lies slightly further to the expected value (about 150~{nm} and 230~{nm}, for Si and diamond, correspondingly). This observation is attributed to the redistribution of radiation between the x and y directions, as shown in Fig.~S2(d) in the SM~\cite{SuppMat}, when the Kerker condition is satisfied. To illustrate this redistribution, we calculated characteristic radiation patterns for silicon nanoparticles with radii of $130\pm10$~nm. This allows us to understand the ``overflowing'' between the magnitudes of $\kappa_{\rm x}$ and $\kappa_{\rm y}$, which becomes pronounced around $R=150~{\rm nm}$.

\paragraph{Radially polarized incident light.} 
For the case of the radially polarized incident light, we obtain the results given in Fig.~\ref{fig:Sec7kappaRadial}. As in the case of circular polarization, due to the axial symmetry of the vector incident field, distributions $\kappa_{\rm x}$ and $\kappa_{\rm y}$ are the same. Besides, we observe remarkable changes in comparison with the incident linear (circular) Gaussian beam. The most evident one is the shift of the radius, which provides the highest forward scattering. The distributions obtained for Si show that the characteristic dips of $\kappa_{\rm i}$ are shifted along the radius axes away from 130-150~{nm} by approximately 100~{nm}. As to the case of diamond, these changes occur beyond the maximal radius we consider in our work and are not shown in Fig.~\ref{fig:Sec7kappaRadial}. As can be seen, the lateral distributions $\kappa_{\rm x,y}$ have no significant changes in the broader range of radii compared to the results obtained for linearly polarized light. Moreover, the distribution of $\kappa_{\rm tot}$ obtained for silicon points to negligible variation over a broad range of radii $R\in [1~{\rm nm}, 190~{\rm nm}]$. These observations might be useful for justification of the dipole approximation when applied to large nanospheres made of silicon or materials with similar refractive indices.

The characteristic shift of the forward scattering condition towards the higher radii is due to excitation of the \emph{anapole} modes~\cite{Wei2016, Krasavin2018, Parker2020}. Due to the superposition of the plane waves at the focus, the excitation of dipoles (electric and magnetic) modes is significantly suppressed, and the resonance of higher-order magnetic and electric modes contributes to the radiation, see Fig.~S2~(c) in the SM~\cite{SuppMat}. For the quadrupole modes, we expect the conditions of significant forward scattering to be satisfied when $a_{_{\rm N=2}}=b_{_{\rm N=2}}$ (see SM~\cite{SuppMat}) and calculations show that it occurs near $R=227~{\rm nm}$ for silicon, and $R=274~{\rm nm}$ for diamond.

So far, we have focused on the computation of the mechanical recoil parameter $\kappa$, central to describing recoil heating effects. While previous studies often approach the recoil heating rate $\Gamma_0$ as a product of $\kappa$ and the scattering cross-section $\sigma_{\rm sc}$~\cite{Seberson2020, Maurer2022, GonzalezBallestero2023}, our methodology isolates $\kappa$ for detailed evaluation. Alongside $\kappa$, our calculation framework naturally determines $\sigma_{\rm sc}$, the results of which are presented in the SM~\cite{SuppMat}, showing significant growth over the given range of radii. The separation of these contributions allows for a more nuanced understanding of the kinematic processes described by $\kappa$, independent of the scattering rate $\nu^{\rm sc}_{_{\rm [ph/s]}}$, which is defined via $\sigma_{\rm sc}$.

The variation of the cross-section ($\sigma_{\rm sc}$) with radius (see Fig.~S5 in the SM~\cite{SuppMat}) dominates over the variation of $\kappa$. As a result, the characteristic dips in the $\kappa_{\rm i}$ contributing to $\Gamma_0$ are not distinctive. Although increasing $\sigma_{\rm sc}$ enhances decoherence and thus degrades experimental sensitivity, its growth can also be viewed in terms of increased temporal sampling density (``signal redundancy''). Since the scattering rate $\nu^{\rm sc}_{{\rm [ph/s]}} \propto I \sigma_{\rm sc}$, a larger $\sigma_{\rm sc}$ leads to more frequent interrogation of the particle’s position. This reduces the acquisition time required for motion tracking and enables operation at a higher feedback bandwidth, provided that the feedback loop is appropriately adjusted. At the same time, the growth of $\sigma_{\rm sc}$ enhances recoil heating and consequently increases $\Gamma_0$, which remains the dominant effect under typical experimental conditions. In this context, $\kappa$ sets the recoil-induced back-action per scattering event, whereas the overall magnitude of $\Gamma_0$ is determined predominantly by the scaling of the total scattering strength. This reflects an inherent trade-off between increased temporal sampling and increased decoherence, as improved temporal resolution comes at the cost of stronger recoil-induced disturbance.

To this extent, we consider the individual, although spatially averaged, scattering event, which highlights the pure kinematic nature of the momentum uncertainty evolution during a photon recoil, excluding the temporal aspect of recoil heating expressed in the rate of such events. Our computational approach provides accurate values for $\kappa$, enabling its independent assessment.

\paragraph*{Final remarks.} 
The presented findings reveal a natural increase in forward scattering, which occurs without the need for auxiliary components and leads to a recoil reduction. We interpret this reduction in recoil as a decrease in back-action at the level of individual scattering events. Our analysis focuses on an optical setup with a single-pass trapping beam, where directivity plays a crucial role. This system represents a particular realization of optical trapping. In contrast, trapping systems based on standing waves are fundamentally different. Since a standing wave forms from two counter-propagating waves, forward and backward directions become indistinguishable, making the concept of the mere use of forward scattering inapplicable. As a result, our findings do not extend to such systems.

The reduction of recoil was considered under idealized conditions, assuming zero absorption. However, in real experiments, one must account for the heating of the trapping object in experimental setups (see, e.g.,~\cite{Millen2014}).
As pointed out in Sec.~4.1 and 4.5 in~\cite{Boyd2020} and~\cite{Devi2023}, the amount of the incident power and the rise of temperature of the trapped object both affect the refractive index of the sphere's material. The effective refractive index $n_{\rm eff}$ with the neglected nonlinear refractive index component can be written in the following form
\begin{equation}
    n_{\rm eff}\approx n +\frac{\dd n }{\dd T} \Delta T\,,
\end{equation}
where $\frac{\dd n }{\dd T}$ is the thermo-optic coefficient and $\Delta T$ is the temperature variation. Such dependence on the temperature assumes altering the conditional radius to obtain the required directivity maxima approaching the Kerker condition. This point means that the choice of the radius of the trapped nanoparticles sample allows for some preliminary corrections in the choice of radii in practice. In the case of silicon, we estimate the thermo-optic coefficient comparing refractive indexes obtained for 300~{K} and 500~{K} (see~\cite{Franta2017, Polyanskiy2024}) and found it to be $(\dd n /\dd T)_{_{\lambda@1064{\rm nm}}} \approx 2.3\cdot 10^{-4}+i\,3.4\cdot 10^{-6}~[\rm K^{-1}]$. The calculations of Eq.~\eqref{eq:CosMean} with this correction coefficient suggest that a silicon nanoparticle illuminated with linearly polarized light and heated up to 1000~{K} shifts the corresponding radius of expected maximal directivity from $\approx130$~{nm} to $125$~{nm}. On the other hand, the estimations of the scattering rate show that the assumption that $\sigma_{\rm sc}/\sigma_{\rm ext}\approx 1$ made in Sec.~\ref{sec:Methods} is expected to hold for the silicon at room conditions, to which we obtain $\sigma_{\rm sc}/\sigma_{\rm ext}\approx 0.999$, weakens for heated material resulting in $\sigma_{\rm sc}/\sigma_{\rm ext}\approx 0.986$ at $R=125~{\rm nm}$ that only would facilitate further heating. These estimations lead to the fact that minimizing the back-action effect might not be easily realized for the optical trapping experiments. However, it can still be applicable for the experiments utilizing the so-called \emph{dark potentials} (see, e.g.,~\cite{Bonvin2024, Dago2024}), where the incident light power can be drastically reduced for only detection needs excluding trapping. At the same time, we acknowledge that our model neglects modifications of the scattering fields caused by absorption heating, which would require a comprehensive and fully dynamical treatment of the refractive index. To partially account for absorption, we retain the factor $\sigma_{\rm sc}/\sigma_{\rm ext}$ in Eq.~(8), which probabilistically includes the fraction of scattered photons. While the static approximation we follow remains valid even for refractive indices with a significant imaginary part, absorption heating and the resulting dynamical changes of the refractive index lie beyond the present scope of this work.

As our final remark, we note that the opportunity to enhance forward scattering has consequences manifested as increased the scattering cross-section. The connection between directivity and scattering cross-section can be understood in terms of the optical theorem, i.e., the growth of amplitude of the forward scattered light, characterized by directivity, leads to the rise of the scattering cross-section.
Thus, within the scope of this work, the following potential avenues for future research can be highlighted: 1) obtaining the extreme directive scattering source with the possibility to vary its radius on demand, and 2) mitigating the dependence of the scattering cross-section on the forward scattered light. The first one might be achieved with further developments in \emph{Mie-tronics} (see, e.g.,~\cite{Kivshar2022, BaratiSedeh2024}) and engineering new scatterers and metamaterials (see, e.g.,~\cite{Shi2022, Parali2024}). As an example, the high directivity of a single nanosphere was obtained by adding extra dielectric shells (dielectric layers) (see, e.g.,~\cite{Liu2012, Liu2014}), or slight deformation of a sphere's shape adding a notch (see, e.g.,~\cite{Krasnok2014}). Concerning the second point, the optical theorem is based on the reciprocity and energy conservation law, weakening the optical theorem would require violation of one of these two. Hence, the perspectives to overcome this correlation can be found by involving non-Hermitian processes.

In the context of exotic scatterer geometries providing high directivity, it is worth noting that although our analysis has been restricted to spherical particles, the approach remains convenient for coated spheres with dielectric shells (see, e.g., Sec.~8.1 in~\cite{Bohren1998} and Sec.~8.2 in~\cite{Tsang2000}) as well as for dielectric spheroids (see, e.g.,~\cite{Asano1975,Ding2023}) under a similar calculation methodology. However, the evaluation of the resulting force from the incident radiation and its consistent relation to an effective definition of $\mathbb{P}_{\rm in}$ remains an issue, particularly for asymmetric objects that may change their orientation within the trap due to wobbling. One may expect that both the extinction coefficient and the scattering characteristics will vary dynamically. At the same time, the computation of the scattered-field amplitudes does not pose a challenge, as it can be performed numerically using established numerical methods (see, e.g.,~\cite{Laing2024}).

\section{Conclusion}\label{sec:summary}
We have discussed the connection between the directivity achieved at scattering from Mie particles and the kinematic recoil component concerning optical levitation. The problem was treated in far-field representation using plane wave decomposition. We have presented the general discussion approach used to obtain the scattered field corresponding to the broad range of focusing NA relevant to tabletop experiments for finding corresponding recoil momenta uncertainty associated with the back-action of an individual scattering event.

For the particular examples of silica, diamond, and silicon nanospheres with diameters up to 500~{nm}, we studied the role of the satisfied Kerker condition in reducing the kinetic energy transferred to the trapped particles. Although we limited ourselves to particular materials, our conclusions may go beyond these particular cases and allow for an understanding of general tendencies for low-absorption materials with a wide range of refractive indices.

For the linearly focused Gaussian beam, the reduction of the total energy transferred to a particle exceeds 40\% in comparison to the linear dipole, and exceeds 80\% when comparing energy acquired for the direction along the beam propagation. The discussed reduction is mitigated by the faster rise of scattering cross-section and, hence, the recoil heating rate growth that causes a faster oscillator decoherence rate, preventing continuous measurement improvement. Nevertheless, the benefit of enhanced directivity remains evident at the level of individual scattering events. The corresponding reduction of momentum uncertainty indicates a substantially weaker disturbance imparted by the probe per interaction.

We have also considered the case of an incident radially polarized trapping beam. Due to the electric and magnetic dipole scattering modes suppression, the Kerker resonance was obtained for the quadrupole components, which required a significant increase of the nanosphere radii to fulfill this condition. For the studied radii range $R\in[1~{\rm nm},250~{\rm nm}]$, the local high directivity was observed only for the silicon nanosphere at 227~nm.

Meeting the Kerker condition does not provide a perfect back-action reduction but a partial one. Nevertheless, showing distinguishable reduction of recoil momentum uncertainty due to naturally approached collinearity between incident and scattered light, our work opens further questions for further optimization approaches aimed at the further suppression of the recoil. Thus, the mechanism of reduction of the recoil should not be overlooked in the context of experimental investigations. Moreover, one may expect that modifying the forward scattering radiation caused by the inherent dielectric properties of the scatterer might be combined with the back-action suppression mechanism based on the self-interference discussed in~\cite{Weiser2025, Gajewski2025}. This combination has the potential to suppress back-action even further exceeding capabilities of each mechanism separately, and we leave this as  an open question for further investigation.

Building on the tendency toward collinearity, the present results may motivate nanophotonic engineering using metamaterials and metastructures to design scatterers with enhanced directivity and near-collinear absorption and emission. Improved control over scattering directivity through such approaches provides a practical route toward reducing recoil-induced disturbance in optically levitated systems. In this limit, the interaction approaches a regime of minimal disturbance, conceptually related to reduced measurement-induced back-action in quantum measurement theory~\cite{Caves1980, Vorontsov1994}. A rigorous quantum-mechanical analysis of this regime, including its possible relevance to quantum nondemolition (QND)~\cite{Braginsky1980, Caves1983} measurement concepts, lies beyond the scope of the present work and remains a direction for future investigation.

\section*{Acknowledgments}
V.S. acknowledges useful discussions and helpful communications with Andrea Aiello, Dmitry Bykov, Vijay Jain, Norbert Lindlein, Andrey Manchev, Florian Marquardt, Dmitriy Pokhabov, Halina Rubinsztein-Dunlop, Colin Sheppard, Markus Sondermann, Sergey Vyatchanin. Technical suggestions on parallel computing from R{\"a}ul Gonz{\'a}lez Garrido and Daniel H{\"a}upl are highly appreciated.

\section*{Data availability statement}
The raw data underlying Fig.~\ref{fig:Sec7kappaLinear}, Fig.~\ref{fig:Sec7kappaRadial}, and Fig.~S5 (Supplementary Materials) are available in the Zenodo repository: \href{https://doi.org/10.5281/zenodo.20021440}{DOI: 10.5281/zenodo.20021440}.

\bibliography{bibtexfile}

@FOOTNOTE{PoyntingVector,
key="PoyntingVector",
note="In the far-field approximation, the momentum of the electromagnetic field is given by $\langle p_{\rm i}\rangle=\frac{1}{c} \int \vec{S}_{\rm i}(\theta,\phi)\, \dd \Omega$, where $\vec{S}_{\rm i}$ is the ${\rm i}$-th component of the Umov–Poynting vector, defined as $\vec{S}(\theta,\phi) = [\vec{E}(\theta,\phi) \cross \vec{H}(\theta,\phi)] = \vec{k}(\theta,\phi)|\vec{E}(\theta,\phi)|^{2}$. This expression follows from the far-field identity $\vec{H}(\theta,\phi) = \vec{k}(\theta,\phi) \cross \vec{E}(\theta,\phi)$. As a result, the averaged momentum becomes $\langle p_{\rm i}\rangle=\frac{1}{c} \int \vec{k}_{\rm i}(\theta,\phi)|\vec{E}(\theta,\phi)|^{2}\, \dd \Omega$, where the probability distribution is proportional to $|\vec{E}(\theta,\phi)|^{2}$, see SM~\cite{SuppMat}."
}

@Article{Asano1975,
  author    = {Shoji Asano and Giichi Yamamoto},
  journal   = {Appl. Opt.},
  title     = {Light Scattering by a Spheroidal Particle},
  year      = {1975},
  month     = {Jan},
  number    = {1},
  pages     = {29--49},
  volume    = {14},
  doi       = {10.1364/AO.14.000029},
  publisher = {Optica Publishing Group},
  url       = {https://opg.optica.org/ao/abstract.cfm?URI=ao-14-1-29},
}

@Article{Laing2024,
  author        = {Shaun Laing and Shelby Klomp and George Winstone and Alexey Grinin and Andrew Dana and Zhiyuan Wang and Kevin Seca Widyatmodjo and James Bateman and Andrew A. Geraci},
  title         = {Optimal displacement detection of arbitrarily-shaped levitated dielectric objects using optical radiation},
  journal="",
  year          = {2024},
  month         = sep,
  archiveprefix = {arXiv},
  eprint        = {2409.00782},
  primaryclass  = {physics.optics},
}

@article{Novotny2017,
  title = {Radiation damping of a polarizable particle},
  author = {Novotny, Lukas},
  journal = {Phys. Rev. A},
  volume = {96},
  issue = {3},
  pages = {032108},
  numpages = {5},
  year = {2017},
  month = {Sep},
  publisher = {American Physical Society},
  doi = {10.1103/PhysRevA.96.032108},
  url = {https://link.aps.org/doi/10.1103/PhysRevA.96.032108}
}

@article{Sinha2022,
doi = {10.1088/1361-6455/ac8efe},
url = {https://dx.doi.org/10.1088/1361-6455/ac8efe},
year = {2022},
month = {oct},
publisher = {IOP Publishing},
volume = {55},
number = {20},
pages = {204002},
author = {Sinha, Kanu and Milonni, Peter W},
title = {Dipoles in blackbody radiation: momentum fluctuations, decoherence, and drag force},
journal = {Journal of Physics B: Atomic, Molecular and Optical Physics},
}

@article{Anishur2021,
  title = {Realizing Einstein's Mirror: Optomechanical Damping with a Thermal Photon Gas},
  author = {Rahman, A. T. M. Anishur and Barker, P. F.},
  journal = {Phys. Rev. Lett.},
  volume = {127},
  issue = {21},
  pages = {213602},
  numpages = {5},
  year = {2021},
  month = {Nov},
  publisher = {American Physical Society},
  doi = {10.1103/PhysRevLett.127.213602},
  url = {https://link.aps.org/doi/10.1103/PhysRevLett.127.213602}
}

@FOOTNOTE{SuppMat,
key="SuppMat",
note="The Supplemental Material includes details on incident and scattering probability distributions in far-field (Sec.~A), on the Kerker effect (Sec.~B), the scattering cross-section (Sec.~C), and a discussion of the tolerance of the approximation employed in this work (Sec.~D).
"
}

@article{Seberson2020,
author = {Seberson, T. and Robicheaux, F.},
doi = {10.1103/PhysRevA.102.033505},
issn = {24699934},
journal = {Physical Review A},
number = {3},
pages = {33505},
publisher = {American Physical Society},
title = {{Distribution of laser shot-noise energy delivered to a levitated nanoparticle}},
url = {https://doi.org/10.1103/PhysRevA.102.033505},
volume = {102},
year = {2020}
}

@book{Hulst1981,
  author = {van de Hulst, H. C.},
  title = {Light scattering by small particles},
  publisher = {Dover Publications},
  year = {1981},
  pages = {470}
}

@book{NovotnyHecht2012,
  author = {Novotny, Lukas},
  title = {Principles of nano-optics},
  publisher = {Cambridge University Press},
  year = {2012},
  pages = {578}
}

@book{Kerker1969,
  author = {Kerker, Milton},
  title = {Scattering of Light and Other Electromagnetic Radiation},
  publisher = {Academic Press},
  year = {1969},
  pages = {666}
}

@book{Tsang2000,
  title={Scattering of electromagnetic waves: theories and applications},
  author={Tsang, Leung and Kong, Jin Au and Ding, Kung-Hau},
  volume={15},
  year={2000},
  publisher={John Wiley \& Sons}
}

@article{OlmosTrigo2020,
  title = {Kerker Conditions upon Lossless, Absorption, and Optical Gain Regimes},
  author = {Olmos-Trigo, Jorge and Sanz-Fern\'andez, Cristina and Abujetas, Diego R. and Lasa-Alonso, Jon and de Sousa, Nuno and Garc\'{\i}a-Etxarri, Aitzol and S\'anchez-Gil, Jos\'e A. and Molina-Terriza, Gabriel and S\'aenz, Juan Jos\'e},
  journal = {Phys. Rev. Lett.},
  volume = {125},
  issue = {7},
  pages = {073205},
  numpages = {6},
  year = {2020},
  month = {Aug},
  publisher = {American Physical Society},
  doi = {10.1103/PhysRevLett.125.073205},
  url = {https://link.aps.org/doi/10.1103/PhysRevLett.125.073205}
}

@Article{Kerker1983,
  author    = {Kerker, M. and Wang, D.-S. and Giles, C. L.},
  journal   = {Journal of the Optical Society of America},
  title     = {Electromagnetic scattering by magnetic spheres},
  year      = {1983},
  issn      = {0030-3941},
  month     = jun,
  number    = {6},
  pages     = {765},
  volume    = {73},
  doi       = {10.1364/josa.73.000765},
  publisher = {Optica Publishing Group},
}

@Article{Jain2016,
  author    = {Vijay Jain and Jan Gieseler and Clemens Moritz and Christoph Dellago and Romain Quidant and Lukas Novotny},
  journal   = {Physical Review Letters},
  title     = {Direct Measurement of Photon Recoil from a Levitated Nanoparticle},
  year      = {2016},
  month     = {jun},
  number    = {24},
  pages     = {243601},
  volume    = {116},
  doi       = {10.1103/physrevlett.116.243601},
  publisher = {American Physical Society ({APS})},
}

@phdthesis{Gieseler2014,
  title = {Dynamics of optically levitated nanoparticles in high vacuum},
  url = {http://dx.doi.org/10.5821/dissertation-2117-95281},
  DOI = {10.5821/dissertation-2117-95281},
  school = {Universitat Politècnica de Catalunya},
  author = {Gieseler,  Jan},
  year = 2014,
  editor = {Quidant,  Romain and Novotny,  Lukas}
}

@Article{Tebbenjohanns2019,
  author    = {Felix Tebbenjohanns and Martin Frimmer and Lukas Novotny},
  journal   = {Physical Review A},
  title     = {Optimal position detection of a dipolar scatterer in a focused field},
  year      = {2019},
  month     = {oct},
  number    = {4},
  pages     = {043821},
  volume    = {100},
  doi       = {10.1103/physreva.100.043821},
  publisher = {American Physical Society ({APS})},
}

@Article{Krylenko2011,
  title={Structure of a laser field of various polarizations in the focal region of an ideal focusing lens. Calculation by methods of scalar diffraction theory},
  author={Krylenko, Yu V and Mikhailov, Yu A and Orekhov, AS and Sklizkov, GV and Chekmarev, AM},
  journal={Journal of Russian Laser Research},
  volume={32},
  pages={19--40},
  year={2011},
  doi       = {10.1007/s10946-011-9186-2},
url={https://doi.org/10.1007/s10946-011-9186-2},
  publisher = {Springer Science and Business Media {LLC}},
}

@Article{Clerk2010,
  author    = {A. A. Clerk and M. H. Devoret and S. M. Girvin and Florian Marquardt and R. J. Schoelkopf},
  journal   = {Reviews of Modern Physics},
  title     = {Introduction to quantum noise, measurement, and amplification},
  year      = {2010},
  month     = {apr},
  number    = {2},
  pages     = {1155--1208},
  volume    = {82},
  doi       = {10.1103/revmodphys.82.1155},
  publisher = {American Physical Society ({APS})},
}

@article{Shi2022,
    author = {Shi, Yuzhi and Song, Qinghua and Toftul, Ivan and Zhu, Tongtong and Yu, Yefeng and Zhu, Weiming and Tsai, Din Ping and Kivshar, Yuri and Liu, Ai Qun},
    title = {Optical manipulation with metamaterial structures},
    journal = {Applied Physics Reviews},
    volume = {9},
    number = {3},
    pages = {031303},
    year = {2022},
    month = {08},
    issn = {1931-9401},
    doi = {10.1063/5.0091280},
    url = {https://doi.org/10.1063/5.0091280}
}

@Book{BornWolf1999,
  author = {Born, Max and Wolf, Emil},
  publisher = {Cambridge University Press},
  title     = {Principles of optics},
  year      = {1999},
  isbn      = {0521642221},
  pages     = {952},
  subtitle  = {electromagnetic theory of propagation, interference and diffraction of light},
}

@Book{Zangwill2013,
  author    = {Zangwill, Andrew},
  publisher = {Cambridge University Press},
  title     = {Modern electrodynamics},
  year      = {2013},
  isbn      = {9780521896979},
  pages     = {977},
}

@Article{Itano1982,
  author    = {Itano, Wayne M. and Wineland, D. J.},
  journal   = {Physical Review A},
  title     = {Laser cooling of ions stored in harmonic and Penning traps},
  year      = {1982},
  issn      = {0556-2791},
  month     = jan,
  number    = {1},
  pages     = {35--54},
  volume    = {25},
  doi       = {10.1103/physreva.25.35},
  publisher = {American Physical Society (APS)},
  url       = {https://link.aps.org/doi/10.1103/PhysRevA.25.35},
}

@Article{Rohrbach2001,
  author    = {Rohrbach, Alexander and Stelzer, Ernst H. K.},
  journal   = {Journal of the Optical Society of America A},
  title     = {Optical trapping of dielectric particles in arbitrary fields},
  year      = {2001},
  issn      = {1520-8532},
  month     = apr,
  number    = {4},
  pages     = {839},
  volume    = {18},
  doi       = {10.1364/josaa.18.000839},
  publisher = {Optica Publishing Group},
}

@Article{Rohrbach2002,
  author    = {Rohrbach, Alexander and Stelzer, Ernst H. K.},
  journal   = {Journal of Applied Physics},
  title     = {Three-dimensional position detection of optically trapped dielectric particles},
  year      = {2002},
  issn      = {1089-7550},
  month     = apr,
  number    = {8},
  pages     = {5474--5488},
  volume    = {91},
  doi       = {10.1063/1.1459748},
  publisher = {AIP Publishing},
}

@Article{Richards1959,
  author               = {Richards, B. and Wolf, E.},
  journal              = {Proceedings of the Royal Society of London. Series A. Mathematical and Physical Sciences},
  title                = {{Electromagnetic Diffraction in Optical Systems. II. Structure of the Image Field in an Aplanatic System}},
  year                 = {1959},
  issn                 = {1471-2946},
  month                = dec,
  number               = {1274},
  pages                = {358--379},
  volume               = {253},
  day                  = {15},
  doi                  = {10.1098/rspa.1959.0200},
  publisher            = {The Royal Society},
  url                  = {http://dx.doi.org/10.1098/rspa.1959.0200},
}

@Article{Toeroek1998,
  author    = {Török, P. and Higdon, P.D. and Juškaitis, R. and Wilson, T.},
  journal   = {Optics Communications},
  title     = {Optimising the image contrast of conventional and confocal optical microscopes imaging finite sized spherical gold scatterers},
  year      = {1998},
  issn      = {0030-4018},
  month     = oct,
  number    = {4–6},
  pages     = {335--341},
  volume    = {155},
  doi       = {10.1016/s0030-4018(98)00384-8},
  publisher = {Elsevier BV},
}

@Article{Lerme2008,
  author    = {Lermé, Jean and Bachelier, Guillaume and Billaud, Pierre and Bonnet, Christophe and Broyer, Michel and Cottancin, Emmanuel and Marhaba, Salem and Pellarin, Michel},
  journal   = {Journal of the Optical Society of America A},
  title     = {Optical response of a single spherical particle in a tightly focused light beam: application to the spatial modulation spectroscopy technique},
  year      = {2008},
  issn      = {1520-8532},
  month     = jan,
  number    = {2},
  pages     = {493},
  volume    = {25},
  doi       = {10.1364/josaa.25.000493},
  publisher = {Optica Publishing Group},
}

@Article{Bekshaev2013,
  author    = {Bekshaev, Aleksandr Y. and Bliokh, Konstantin Y. and Nori, Franco},
  journal   = {Optics Express},
  title     = {Mie scattering and optical forces from evanescent fields: A complex-angle approach},
  year      = {2013},
  issn      = {1094-4087},
  month     = mar,
  number    = {6},
  pages     = {7082},
  volume    = {21},
  doi       = {10.1364/oe.21.007082},
  publisher = {Optica Publishing Group},
}

@Article{Li2005,
  author    = {Li, Peng and Shi, Kebin and Liu, Zhiwen},
  journal   = {Optics Express},
  title     = {Optical scattering spectroscopy by using tightly focused supercontinuum},
  year      = {2005},
  issn      = {1094-4087},
  number    = {22},
  pages     = {9039},
  volume    = {13},
  doi       = {10.1364/opex.13.009039},
  publisher = {Optica Publishing Group},
}

@Article{Volpe2007,
  author   = {Volpe, Giovanni and Kozyreff, Gregory and Petrov, Dmitri},
  journal  = {Journal of Applied Physics},
  title    = {Backscattering position detection for photonic force microscopy},
  year     = {2007},
  issn     = {0021-8979},
  month    = {10},
  number   = {8},
  pages    = {084701},
  volume   = {102},
  doi      = {10.1063/1.2799047},
  url      = {https://doi.org/10.1063/1.2799047},
}

@Article{Ding2023,
  author    = {Ding, Jiachen and Yang, Ping},
  journal   = {Optics Express},
  title     = {Lorenz-Mie theory-type solution for light scattering by spheroids with small-to-large size parameters and aspect ratios},
  year      = {2023},
  issn      = {1094-4087},
  month     = nov,
  number    = {24},
  pages     = {40937},
  volume    = {31},
  doi       = {10.1364/oe.505416},
  publisher = {Optica Publishing Group},
}

@Article{Gouesbet2015,
  author    = {Gouesbet, Gérard and Lock, James A.},
  journal   = {Journal of Quantitative Spectroscopy and Radiative Transfer},
  title     = {On the electromagnetic scattering of arbitrary shaped beams by arbitrary shaped particles: A review},
  year      = {2015},
  issn      = {0022-4073},
  month     = sep,
  pages     = {31--49},
  volume    = {162},
  doi       = {10.1016/j.jqsrt.2014.11.017},
  publisher = {Elsevier BV},
}

@Article{Lamberg2023,
  author    = {Lamberg, Joel and Zarrinkhat, Faezeh and Tamminen, Aleksi and Ala-Laurinaho, Juha and Rius, Juan and Romeu, Jordi and Khaled, Elsayed E. M. and Taylor, Zachary},
  journal   = {Optics Express},
  title     = {Mie scattering with 3D angular spectrum method},
  year      = {2023},
  issn      = {1094-4087},
  month     = oct,
  number    = {23},
  pages     = {38653},
  volume    = {31},
  doi       = {10.1364/oe.504791},
  publisher = {Optica Publishing Group},
}

@Article{RanhaNeves2019,
  author    = {Ranha Neves, Antonio Alvaro and Cesar, Carlos Lenz},
  journal   = {Journal of the Optical Society of America B},
  title     = {Analytical calculation of optical forces on spherical particles in optical tweezers: tutorial},
  year      = {2019},
  issn      = {1520-8540},
  month     = may,
  number    = {6},
  pages     = {1525},
  volume    = {36},
  doi       = {10.1364/josab.36.001525},
  publisher = {Optica Publishing Group},
}

@Article{Kimble2001,
  author    = {Kimble, H. J. and Levin, Yuri and Matsko, Andrey B. and Thorne, Kip S. and Vyatchanin, Sergey P.},
  journal   = {Physical Review D},
  title     = {Conversion of conventional gravitational-wave interferometers into quantum nondemolition interferometers by modifying their input and/or output optics},
  year      = {2001},
  issn      = {1089-4918},
  month     = dec,
  number    = {2},
  pages     = {022002},
  volume    = {65},
  doi       = {10.1103/physrevd.65.022002},
  publisher = {American Physical Society (APS)},
}

@Book{Bohren1998,
  author    = {Bohren, Craig F. and Huffman, Donald R.},
  publisher = {Wiley-Interscience},
  title     = {Absorption and Scattering of Light by Small Particles (Wiley Science Paperback Series)},
  year      = {1998},
  isbn      = {9780471293408},
  pages     = {544},
}

@Article{Davidson2004,
  author    = {Davidson, Nir and Bokor, Nándor},
  journal   = {Optics Letters},
  title     = {High-numerical-aperture focusing of radially polarized doughnut beams with a parabolic mirror and a flat diffractive lens},
  year      = {2004},
  issn      = {1539-4794},
  month     = jun,
  number    = {12},
  pages     = {1318},
  volume    = {29},
  doi       = {10.1364/ol.29.001318},
  publisher = {Optica Publishing Group},
}

@Article{GonzalezBallestero2019,
  author    = {Gonzalez-Ballestero, C. and Maurer, P. and Windey, D. and Novotny, L. and Reimann, R. and Romero-Isart, O.},
  journal   = {Physical Review A},
  title     = {Theory for cavity cooling of levitated nanoparticles via coherent scattering: Master equation approach},
  year      = {2019},
  issn      = {2469-9934},
  month     = jul,
  number    = {1},
  pages     = {013805},
  volume    = {100},
  doi       = {10.1103/physreva.100.013805},
  publisher = {American Physical Society (APS)},
}

@Article{Parali2024,
  author    = {Paral{\i}, Ufuk and {\"U}st{\"u}n, Kadir and Giden, Ibrahim Halil},
  journal   = {Scientific Reports},
  title     = {Enhancement of optical levitation with hyperbolic metamaterials},
  volume    = {14},
  number    = {1},
  pages     = {1734},
  year      = {2024},
  doi       = {10.1038/s41598-024-51284-4},
  publisher = {Nature Publishing Group UK London}
}

@Article{Wang2025,
  author    = {Wang, Long and Zhou, Lei-Ming and Tian, Yuan and Liu, Lyu-Hang and Guo, Guang-Can and Zheng, Yu and Sun, Fang-Wen},
  journal   = {Journal of the Optical Society of America B},
  title     = {Optimal position detection of an optically levitated Mie particle},
  year      = {2025},
  issn      = {1520-8540},
  month     = feb,
  number    = {3},
  pages     = {645},
  volume    = {42},
  doi       = {10.1364/josab.544546},
  publisher = {Optica Publishing Group},
}

@Article{GonzalezBallestero2023,
  author    = {Gonzalez-Ballestero, C. and Zielińska, J.A. and Rossi, M. and Militaru, A. and Frimmer, M. and Novotny, L. and Maurer, P. and Romero-Isart, O.},
  journal   = {PRX Quantum},
  title     = {Suppressing Recoil Heating in Levitated Optomechanics Using Squeezed Light},
  year      = {2023},
  issn      = {2691-3399},
  month     = sep,
  number    = {3},
  pages     = {030331},
  volume    = {4},
  doi       = {10.1103/prxquantum.4.030331},
  publisher = {American Physical Society (APS)},
}

@Article{Braginsky1980,
  author    = {Braginsky, Vladimir B. and Vorontsov, Yuri I. and Thorne, Kip S.},
  journal   = {Science},
  title     = {Quantum Nondemolition Measurements},
  year      = {1980},
  issn      = {1095-9203},
  month     = aug,
  number    = {4456},
  pages     = {547--557},
  volume    = {209},
  doi       = {10.1126/science.209.4456.547},
  publisher = {American Association for the Advancement of Science (AAAS)},
}

@Article{Pesce2020,
  author    = {Pesce, Giuseppe and Jones, Philip H. and Maragò, Onofrio M. and Volpe, Giovanni},
  journal   = {The European Physical Journal Plus},
  title     = {Optical tweezers: theory and practice},
  year      = {2020},
  issn      = {2190-5444},
  month     = dec,
  number    = {12},
  volume    = {135},
  pages     = {949},
  url       = {https://link.springer.com/article/10.1140/epjp/s13360-020-00843-5},
  publisher = {Springer Science and Business Media LLC},
}

@PhdThesis{Jain2017,
  author    = {{Jain, Vijay}},
  title     = {Levitated optomechanics at the photon recoil limit},
  year      = {2017},
  copyright = {http://rightsstatements.org/page/InC-NC/1.0/},
  doi       = {10.3929/ETHZ-B-000200312},
  school    = {ETHZ},
  publisher = {ETH Zurich},
}

@Article{Gajewski2025,
  title = {Backaction suppression in levitated optomechanics using reflective boundaries},
  author = {Gajewski, Rafa\l{} and Bateman, James},
  journal = {Phys. Rev. Res.},
  volume = {7},
  issue = {2},
  pages = {023041},
  numpages = {9},
  year = {2025},
  month = {Apr},
  publisher = {American Physical Society},
  doi = {10.1103/PhysRevResearch.7.023041},
  url = {https://link.aps.org/doi/10.1103/PhysRevResearch.7.023041}
}

@PhdThesis{GajewskiTH2024,
  author = {Gajewski, Rafal},
  school = {Swansea University},
  title  = {Backaction suppression in levitated optomechanics},
  doi    = {10.23889/suthesis.67075},
}

@Article{Abbassi2024,
   title = {Derivation of recoil heating of a Rayleigh scatterer from the quantum fluctuations of the electromagnetic fields},
  author = {Abbassi, Mohammad Ali},
  journal = {Phys. Rev. A},
  volume = {110},
  issue = {5},
  pages = {053503},
  numpages = {10},
  year = {2024},
  month = {Nov},
  publisher = {American Physical Society},
  doi = {10.1103/PhysRevA.110.053503},
  url = {https://link.aps.org/doi/10.1103/PhysRevA.110.053503}
}

@Article{Vorontsov1994,
  author    = {Vorontsov, Yurii I.},
  journal   = {Uspekhi Fizicheskih Nauk},
  title     = {Standard quantum limits of measurement error and methods of overcoming them},
  year      = {1994},
  issn      = {1996-6652},
  number    = {1},
  pages     = {89},
  volume    = {164},
  doi       = {10.3367/ufnr.0164.199401d.0089},
  publisher = {Uspekhi Fizicheskikh Nauk (UFN) Journal},
}

@InBook{Caves1983,
  author    = {Caves, Carlton M.},
  pages     = {567--626},
  publisher = {Springer US},
  title     = {Quantum Nondemolition Measurements},
  year      = {1983},
  isbn      = {9781461337126},
  booktitle = {Quantum Optics, Experimental Gravity, and Measurement Theory},
  doi       = {10.1007/978-1-4613-3712-6_24},
}

@Article{Liu2018,
  author    = {Liu, Wei and Kivshar, Yuri S.},
  journal   = {Optics Express},
  title     = {Generalized {K}erker effects in nanophotonics and meta-optics [Invited]},
  year      = {2018},
  issn      = {1094-4087},
  month     = may,
  number    = {10},
  pages     = {13085},
  volume    = {26},
  doi       = {10.1364/oe.26.013085},
  publisher = {Optica Publishing Group},
}

@Article{Liu2012,
  author    = {Liu, Wei and Miroshnichenko, Andrey E. and Neshev, Dragomir N. and Kivshar, Yuri S.},
  journal   = {ACS Nano},
  title     = {Broadband Unidirectional Scattering by Magneto-Electric Core–Shell Nanoparticles},
  year      = {2012},
  issn      = {1936-086X},
  month     = may,
  number    = {6},
  pages     = {5489--5497},
  volume    = {6},
  doi       = {10.1021/nn301398a},
  publisher = {American Chemical Society (ACS)},
}

@Article{Krasnok2014,
  author    = {Krasnok, Alexander E. and Simovski, Constantin R. and Belov, Pavel A. and Kivshar, Yuri S.},
  journal   = {Nanoscale},
  title     = {Superdirective dielectric nanoantennas},
  year      = {2014},
  issn      = {2040-3372},
  number    = {13},
  pages     = {7354--7361},
  volume    = {6},
  doi       = {10.1039/c4nr01231c},
  publisher = {Royal Society of Chemistry (RSC)},
}

@Article{Nechayev2019,
  author    = {Nechayev, Sergey and Eismann, Jörg S. and Neugebauer, Martin and Woźniak, Paweł and Bag, Ankan and Leuchs, Gerd and Banzer, Peter},
  journal   = {Physical Review A},
  title     = {Huygens’ dipole for polarization-controlled nanoscale light routing},
  year      = {2019},
  issn      = {2469-9934},
  month     = apr,
  number    = {4},
  pages     = {041801},
  volume    = {99},
  doi       = {10.1103/physreva.99.041801},
  publisher = {American Physical Society (APS)},
}

@Article{Liu2014,
  author    = {Liu, Wei and Zhang, Jianfa and Lei, Bing and Ma, Haotong and Xie, Wenke and Hu, Haojun},
  journal   = {Optics Express},
  title     = {Ultra-directional forward scattering by individual core-shell nanoparticles},
  year      = {2014},
  issn      = {1094-4087},
  month     = jun,
  number    = {13},
  pages     = {16178},
  volume    = {22},
  doi       = {10.1364/oe.22.016178},
  publisher = {Optica Publishing Group},
}

@Article{NietoVesperinas2010,
  author    = {Nieto-Vesperinas, M. and Gomez-Medina, R. and Saenz, J. J.},
  journal   = {Journal of the Optical Society of America A},
  title     = {Angle-suppressed scattering and optical forces on submicrometer dielectric particles},
  year      = {2010},
  issn      = {1520-8532},
  month     = dec,
  number    = {1},
  pages     = {54},
  volume    = {28},
  doi       = {10.1364/josaa.28.000054},
  publisher = {Optica Publishing Group},
}

@Article{Caves1980,
  author    = {Caves, Carlton M. and Thorne, Kip S. and Drever, Ronald W. P. and Sandberg, Vernon D. and Zimmermann, Mark},
  journal   = {Reviews of Modern Physics},
  title     = {On the measurement of a weak classical force coupled to a quantum-mechanical oscillator. {I}. {I}ssues of principle},
  year      = {1980},
  issn      = {0034-6861},
  month     = apr,
  number    = {2},
  pages     = {341--392},
  volume    = {52},
  doi       = {10.1103/revmodphys.52.341},
  publisher = {American Physical Society (APS)},
}

@techreport{Maetzler2002,
  author    = {Mätzler, Christian},
  title     = {MATLAB Functions for Mie Scattering and Absorption, Version 1},
  year      = {2002},
  doi       = {10.7892/BORIS.146551},
  institution ={Institute of Applied Physics, University of Bern},
}

@Article{Rohrbach2004,
  author    = {Rohrbach, Alexander and Kress, Holger and Stelzer, Ernst H. K.},
  journal   = {Applied Optics},
  title     = {Reply to comment on “{T}rapping force, force constant, and potential depths for dielectric spheres in the presence of spherical aberrations”},
  year      = {2004},
  issn      = {1539-4522},
  month     = mar,
  number    = {9},
  pages     = {1827},
  volume    = {43},
  doi       = {10.1364/ao.43.001827},
  publisher = {Optica Publishing Group},
}

@Article{Rohrbach2005,
  author    = {Rohrbach, Alexander},
  journal   = {Physical Review Letters},
  title     = {Stiffness of Optical Traps: Quantitative Agreement between Experiment and Electromagnetic Theory},
  year      = {2005},
  issn      = {1079-7114},
  month     = oct,
  number    = {16},
  pages     = {168102},
  volume    = {95},
  doi       = {10.1103/physrevlett.95.168102},
  publisher = {American Physical Society (APS)},
}

@Article{Salakhutdinov2020,
  author    = {Salakhutdinov, Vsevolod and Sondermann, Markus and Carbone, Luigi and Giacobino, Elisabeth and Bramati, Alberto and Leuchs, Gerd},
  journal   = {Physical Review Letters},
  title     = {Single Photons Emitted by Nanocrystals Optically Trapped in a Deep Parabolic Mirror},
  year      = {2020},
  issn      = {1079-7114},
  month     = jan,
  number    = {1},
  pages     = {013607},
  volume    = {124},
  doi       = {10.1103/physrevlett.124.013607},
  publisher = {American Physical Society (APS)},
}

@Article{Ahn2018,
  author    = {Ahn, Jonghoon and Xu, Zhujing and Bang, Jaehoon and Deng, Yu-Hao and Hoang, Thai M. and Han, Qinkai and Ma, Ren-Min and Li, Tongcang},
  journal   = {Physical Review Letters},
  title     = {Optically Levitated Nanodumbbell Torsion Balance and GHz Nanomechanical Rotor},
  year      = {2018},
  issn      = {1079-7114},
  month     = jul,
  number    = {3},
  pages     = {033603},
  volume    = {121},
  doi       = {10.1103/physrevlett.121.033603},
  publisher = {American Physical Society (APS)},
}

@Article{Bang2020,
  author    = {Bang, Jaehoon and Seberson, T. and Ju, Peng and Ahn, Jonghoon and Xu, Zhujing and Gao, Xingyu and Robicheaux, F. and Li, Tongcang},
  journal   = {Physical Review Research},
  title     = {Five-dimensional cooling and nonlinear dynamics of an optically levitated nanodumbbell},
  year      = {2020},
  issn      = {2643-1564},
  month     = oct,
  number    = {4},
  pages     = {043054},
  volume    = {2},
  doi       = {10.1103/physrevresearch.2.043054},
  publisher = {American Physical Society (APS)},
}

@Article{Gosling2024,
  author        = {Jonathan M. H. Gosling and Markus Rademacher and Jence T. Mulder and Arjan J. Houtepen and Marko Toroš and A. T. M. Anishur Rahman and Antonio Pontin and P. F. Barker},
  title         = {Levitodynamic spectroscopy for single nanoparticle characterisation},
  year          = {2024},
  month         = jan,
  archiveprefix = {arXiv},
  journal       = "",
  eprint        = {2401.11551},
  primaryclass  = {physics.optics},
}

@Article{Kuhn2017,
  author    = {Kuhn, Stefan and Kosloff, Alon and Stickler, Benjamin A. and Patolsky, Fernando and Hornberger, Klaus and Arndt, Markus and Millen, James},
  journal   = {Optica},
  title     = {Full rotational control of levitated silicon nanorods},
  year      = {2017},
  issn      = {2334-2536},
  month     = mar,
  number    = {3},
  pages     = {356},
  volume    = {4},
  doi       = {10.1364/optica.4.000356},
  publisher = {Optica Publishing Group},
}

@Article{Feng2001,
  author    = {Feng, Simin and Winful, Herbert G.},
  journal   = {Optics Letters},
  title     = {Physical origin of the {G}ouy phase shift},
  year      = {2001},
  issn      = {1539-4794},
  month     = apr,
  number    = {8},
  pages     = {485},
  volume    = {26},
  doi       = {10.1364/ol.26.000485},
  publisher = {Optica Publishing Group},
}

@Article{Franta2017,
  author    = {Franta, Daniel and Dubroka, Adam and Wang, Chennan and Giglia, Angelo and Vohánka, Jirí and Franta, Pavel and Ohlídal, Ivan},
  journal   = {Applied Surface Science},
  title     = {Temperature-dependent dispersion model of float zone crystalline silicon},
  year      = {2017},
  issn      = {0169-4332},
  month     = nov,
  pages     = {405--419},
  volume    = {421},
  doi       = {10.1016/j.apsusc.2017.02.021},
  publisher = {Elsevier BV},
}

@Article{Polyanskiy2024,
   title={Refractiveindex. info database of optical constants},
  author={Polyanskiy, Mikhail N},
  journal={Scientific Data},
  volume={11},
  number={1},
  pages={94},
  year={2024},
  publisher={Nature Publishing Group UK London},
  doi       = {10.1038/s41597-023-02898-2},
  url       = {https://www.nature.com/articles/s41597-023-02898-2}
}

@Book{Boyd2020,
  author    = {Boyd, Robert W.},
  publisher = {Elsevier Science and Technology},
  title     = {Nonlinear Optics},
  year      = {2020},
  isbn      = {9780128110027},
  pages={634}
}

@Article{Devi2023,
  author    = {Devi, Anita and Neupane, Krishna and Jung, Haksung and Neuman, Keir C. and Woodside, Michael T.},
  journal   = {Biophysical Journal},
  title     = {Nonlinear effects in optical trapping of titanium dioxide and diamond nanoparticles},
  year      = {2023},
  issn      = {0006-3495},
  month     = sep,
  number    = {17},
  pages     = {3439--3446},
  volume    = {122},
  doi       = {10.1016/j.bpj.2023.07.018},
  publisher = {Elsevier BV},
}

@article{Dago2024,
author = {Salamb\^{o} Dago and J. Rieser and M. A. Ciampini and V. Mlyn\'{a}\v{r} and A. Kugi and M. Aspelmeyer and A. Deutschmann-Olek and N. Kiesel},
journal = {Opt. Express},
number = {25},
pages = {45133--45141},
publisher = {Optica Publishing Group},
title = {Stabilizing nanoparticles in the intensity minimum: feedback levitation on an inverted potential},
volume = {32},
month = {Dec},
year = {2024},
url = {https://opg.optica.org/oe/abstract.cfm?URI=oe-32-25-45133},
doi = {10.1364/OE.541267}
}

@Article{Bonvin2024,
  author    = {Bonvin, Eric and Devaud, Louisiane and Rossi, Massimiliano and Militaru, Andrei and Dania, Lorenzo and Bykov, Dmitry S. and Teller, Markus and Northup, Tracy E. and Novotny, Lukas and Frimmer, Martin},
  journal   = {Physical Review Research},
  title     = {Hybrid {P}aul-optical trap with large optical access for levitated optomechanics},
  year      = {2024},
  issn      = {2643-1564},
  month     = nov,
  number    = {4},
  pages     = {043129},
  volume    = {6},
  doi       = {10.1103/physrevresearch.6.043129},
  publisher = {American Physical Society (APS)},
}

@Article{Wei2016,
  author    = {Wei, Lei and Xi, Zheng and Bhattacharya, Nandini and Urbach, H. Paul},
  journal   = {Optica},
  title     = {Excitation of the radiationless anapole mode},
  year      = {2016},
  issn      = {2334-2536},
  month     = jul,
  number    = {8},
  pages     = {799},
  volume    = {3},
  doi       = {10.1364/optica.3.000799},
  publisher = {Optica Publishing Group},
}

@Article{Krasavin2018,
  title={Generalization of the optical theorem: experimental proof for radially polarized beams},
  author={Krasavin, Alexey V and Segovia, Paulina and Dubrovka, Rostyslav and Olivier, Nicolas and Wurtz, Gregory A and Ginzburg, Pavel and Zayats, Anatoly V},
  journal={Light: Science \& Applications},
  volume={7},
  number={1},
  pages={36},
  year={2018},
  doi       = {10.1038/s41377-018-0025-x},
url={https://doi.org/10.1038/s41377-018-0025-x},
  publisher={Nature Publishing Group UK London}
}

@Article{Parker2020,
  author    =  {Parker, John A. and Sugimoto, Hiroshi and Coe, Brighton and Eggena, Daniel and Fujii, Minoru and Scherer, Norbert F. and Gray, Stephen K. and Manna, Uttam},
  journal   = {Physical Review Letters},
  title     = {Excitation of Nonradiating Anapoles in Dielectric Nanospheres},
  year      = {2020},
  issn      = {1079-7114},
  month     = mar,
  number    = {9},
  pages     = {097402},
  volume    = {124},
  doi       = {10.1103/physrevlett.124.097402},
  publisher = {American Physical Society (APS)},
}

@Article{Millen2014,
  author    = {Millen, J. and Deesuwan, T. and Barker, P. and Anders, J.},
  journal   = {Nature Nanotechnology},
  title     = {Nanoscale temperature measurements using non-equilibrium Brownian dynamics of a levitated nanosphere},
  year      = {2014},
  issn      = {1748-3395},
  month     = may,
  number    = {6},
  pages     = {425--429},
  volume    = {9},
  doi       = {10.1038/nnano.2014.82},
  publisher = {Springer Science and Business Media LLC},
}

@Article{Kivshar2022,
  author    = {Kivshar, Yuri},
  journal   = {Nano Letters},
  title     = {The Rise of {M}ie-tronics},
  year      = {2022},
  issn      = {1530-6992},
  month     = apr,
  number    = {9},
  pages     = {3513--3515},
  volume    = {22},
  doi       = {10.1021/acs.nanolett.2c00548},
  publisher = {American Chemical Society (ACS)},
}

@Article{BaratiSedeh2024,
  author    = {Barati Sedeh, Hooman and Litchinitser, Natalia M.},
  journal   = {Photonics Research},
  title     = {From non-scattering to super-scattering with {M}ie-tronics},
  year      = {2024},
  issn      = {2327-9125},
  month     = mar,
  number    = {4},
  pages     = {608},
  volume    = {12},
  doi       = {10.1364/prj.503182},
  publisher = {Optica Publishing Group},
}

@Article{Volpe2023,
  author    = {Volpe et al., Giovanni},
  journal   = {Journal of Physics: Photonics},
  title     = {Roadmap for optical tweezers},
  year      = {2023},
  issn      = {2515-7647},
  month     = apr,
  number    = {2},
  pages     = {022501},
  volume    = {5},
  doi       = {10.1088/2515-7647/acb57b},
  publisher = {IOP Publishing},
}

@Article{Winstone2023,
  author        = {George Winstone and Alexey Grinin and Mishkat Bhattacharya and Andrew A. Geraci and Tongcang Li and Peter J. Pauzauskie and Nick Vamivakas},
  title         = {Optomechanics of optically-levitated particles: A tutorial and perspective},
  year          = {2023},
  month         = jul,
  archiveprefix = {arXiv},
  journal       = "",
  eprint        = {2307.11858},
  primaryclass  = {quant-ph},
}

@article{Millen2020,
doi = {10.1088/1361-6633/ab6100},
url = {https://dx.doi.org/10.1088/1361-6633/ab6100},
year = {2020},
month = {jan},
publisher = {IOP Publishing},
volume = {83},
number = {2},
pages = {026401},
author = {Millen, James and Monteiro, Tania S and Pettit, Robert and Vamivakas, A Nick},
title = {Optomechanics with levitated particles},
journal = {Reports on Progress in Physics},
}

@Article{Jin2024,
  author        = {Yuanbin Jin and Kunhong Shen and Peng Ju and Tongcang Li},
  title         = {Towards real-world applications of levitated optomechanics},
  year          = {2024},
  month         = jul,
  journal       = "",
  archiveprefix = {arXiv},
  eprint        = {2407.12496},
  primaryclass  = {physics.optics},
}

@Article{Gieseler2021,
  author    = {Gieseler, Jan and Gomez-Solano, Juan Ruben and Magazzù, Alessandro and Pérez Castillo, Isaac and Perez Garcia, Laura and Gironella-Torrent, Marta and Viader-Godoy, Xavier and Ritort, Felix and Pesce, Giuseppe and Arzola, Alejandro V. and Volke-Sepulveda, Karen and Volpe, Giovanni},
  journal   = {Advances in Optics and Photonics},
  title     = {Optical tweezers — from calibration to applications: a tutorial},
  year      = {2021},
  issn      = {1943-8206},
  month     = mar,
  number    = {1},
  pages     = {74},
  volume    = {13},
  doi       = {10.1364/aop.394888},
  publisher = {Optica Publishing Group},
}

@Article{GonzalezBallestero2021,
author = {C. Gonzalez-Ballestero  and M. Aspelmeyer  and L. Novotny  and R. Quidant  and O. Romero-Isart },
title = {Levitodynamics: Levitation and control of microscopic objects in vacuum},
journal = {Science},
volume = {374},
number = {6564},
pages = {eabg3027},
year = {2021},
doi = {10.1126/science.abg3027},
URL = {https://www.science.org/doi/abs/10.1126/science.abg3027}
}

@Article{Lepeshov2023,
  author    = {Lepeshov, Sergei and Meyer, Nadine and Maurer, Patrick and Romero-Isart, Oriol and Quidant, Romain},
  journal   = {Physical Review Letters},
  title     = {Levitated Optomechanics with Meta-Atoms},
  year      = {2023},
  issn      = {1079-7114},
  month     = jun,
  number    = {23},
  pages     = {233601},
  volume    = {130},
  doi       = {10.1103/physrevlett.130.233601},
  publisher = {American Physical Society (APS)},
}

@Article{Mao2024,
  author    = {Mao, Libang and Toftul, Ivan and Balendhran, Sivacarendran and Taha, Mohammad and Kivshar, Yuri and Kruk, Sergey},
  journal   = {Laser \& Photonics Reviews},
  title     = {Switchable Optical Trapping of Mie‐Resonant Phase‐Change Nanoparticles},
  year      = {2024},
  pages     = {2400767},
  issn      = {1863-8899},
  month     = oct,
  doi       = {10.1002/lpor.202400767},
  publisher = {Wiley},
}

@Article{Weiser2025,
  author    = {Weiser, Y. and Faorlin, T. and Panzl, L. and Lafenthaler, T. and Dania, L. and Bykov, D. S. and Monz, T. and Blatt, R. and Cerchiari, G.},
  journal   = {Physical Review A},
  title     = {Backaction suppression for levitated dipolar scatterers},
  year      = {2025},
  issn      = {2469-9934},
  month     = jan,
  number    = {1},
  pages     = {013503},
  volume    = {111},
  doi       = {10.1103/physreva.111.013503},
  publisher = {American Physical Society (APS)},
}

@Article{Stilgoe2008,
  author    = {Stilgoe, Alexander B. and Nieminen, Timo A. and Knöener, Gregor and Heckenberg, Norman R. and Rubinsztein-Dunlop, Halina},
  journal   = {Optics Express},
  title     = {The effect of {M}ie resonances on trapping in optical tweezers},
  year      = {2008},
  issn      = {1094-4087},
  month     = sep,
  number    = {19},
  pages     = {15039},
  volume    = {16},
  doi       = {10.1364/oe.16.015039},
  publisher = {Optica Publishing Group},
}

@Article{Shilkin2022,
  author    = {Shilkin, D. A. and Fedyanin, A. A.},
  journal   = {JETP Letters},
  title     = {Optical Levitation of {M}ie-Resonant Silicon Particles in the Field of Bloch Surface Electromagnetic Waves},
  year      = {2022},
  issn      = {1090-6487},
  month     = feb,
  number    = {3},
  pages     = {136--140},
  volume    = {115},
  doi       = {10.1134/s0021364022030092},
  publisher = {Pleiades Publishing Ltd},
}

@Article{Monteiro2020,
  author    = {Monteiro, Fernando and Afek, Gadi and Carney, Daniel and Krnjaic, Gordan and Wang, Jiaxiang and Moore, David C.},
  journal   = {Physical Review Letters},
  title     = {Search for Composite Dark Matter with Optically Levitated Sensors},
  year      = {2020},
  issn      = {1079-7114},
  month     = oct,
  number    = {18},
  pages     = {181102},
  volume    = {125},
  doi       = {10.1103/physrevlett.125.181102},
  publisher = {American Physical Society (APS)},
}

@Article{Kalia2024,
  author    = {Kalia, Saarik and Budker, Dmitry and Kimball, Derek F. Jackson and Ji, Wei and Liu, Zhen and Sushkov, Alexander O. and Timberlake, Chris and Ulbricht, Hendrik and Vinante, Andrea and Wang, Tao},
  journal   = {Physical Review D},
  title     = {Ultralight dark matter detection with levitated ferromagnets},
  year      = {2024},
  month     = {Dec},
  pages     = {115029},
  volume    = {110},
  doi       = {10.1103/PhysRevD.110.115029},
  issue     = {11},
  numpages  = {21},
  publisher = {American Physical Society},
  url       = {https://link.aps.org/doi/10.1103/PhysRevD.110.115029},
}

@Article{Kilian2024,
  author   = {Kilian, Eva and Rademacher, Markus and Gosling, Jonathan M. H. and Iacoponi, Julian H. and Alder, Fiona and Toroš, Marko and Pontin, Antonio and Ghag, Chamkaur and Bose, Sougato and Monteiro, Tania S. and Barker, P. F.},
  journal  = {AVS Quantum Science},
  title    = {Dark matter searches with levitated sensors},
  year     = {2024},
  issn     = {2639-0213},
  month    = {09},
  number   = {3},
  pages    = {030503},
  volume   = {6},
  doi      = {10.1116/5.0200916},
  url      = {https://doi.org/10.1116/5.0200916}
}

@article{Winkler2024,
  author = {Klemens Winkler and Anton V. Zasedatelev and Benjamin A. Stickler and Uroš Delić and Andreas Deutschmann-Olek and Markus Aspelmeyer},
  title = {Steady-state entanglement of interacting masses in free space through optimal feedback control},
  journal = {Phys. Rev. Res.},
  volume = {7},
  issue = {4},
  pages = {043298},
  numpages = {10},
  year = {2025},
  month = {Dec},
  publisher = {American Physical Society},
  doi = {10.1103/hxc8-fxcb},
  url = {https://link.aps.org/doi/10.1103/hxc8-fxcb}
}

@Article{Zambon2024,
  author        = {Nicola Carlon Zambon and Massimiliano Rossi and Martin Frimmer and Lukas Novotny and Carlos Gonzalez-Ballestero and Oriol Romero-Isart and Andrei Militaru},
  title         = {Motional entanglement of remote optically levitated nanoparticles},
  year          = {2024},
  month         = aug,
  journal       = "",
  archiveprefix = {arXiv},
  eprint        = {2408.14439},
  primaryclass  = {quant-ph},
}

@Article{Poddubny2024,
  author        = {Alexander N. Poddubny and Klemens Winkler and Benjamin A. Stickler and Uros Delic and Markus Aspelmeyer and Anton V. Zasedatelev},
  title         = {Nonequilibrium entanglement between levitated masses under optimal control},
  year          = {2024},
  month         = aug,
  journal       = "",
  archiveprefix = {arXiv},
  eprint        = {2408.06251},
  primaryclass  = {quant-ph},
}

@Article{Vinante2019,
  author    = {Vinante, A. and Pontin, A. and Rashid, M. and Toros, M. and Barker, P. F. and Ulbricht, H.},
  journal   = {Physical Review A},
  title     = {Testing collapse models with levitated nanoparticles: Detection challenge},
  year      = {2019},
  issn      = {2469-9934},
  month     = jul,
  number    = {1},
  pages     = {012119},
  volume    = {100},
  doi       = {10.1103/physreva.100.012119},
  publisher = {American Physical Society (APS)},
}

@Article{Kivshar2017,
  author    = {Kivshar, Yuri and Miroshnichenko, Andrey},
  journal   = {Optics and Photonics News},
  title     = {Meta-Optics with Mie Resonances},
  year      = {2017},
  issn      = {1541-3721},
  month     = jan,
  number    = {1},
  pages     = {24},
  volume    = {28},
  doi       = {10.1364/opn.28.1.000024},
  publisher = {The Optical Society},
}

@Article{Maurer2022,
  author    = {Maurer, Patrick and Gonzalez-Ballestero, Carlos and Romero-Isart, Oriol},
  journal   = {Phys. Rev. A},
  title     = {Quantum theory of light interaction with a {L}orenz-{M}ie particle: Optical detection and three-dimensional ground-state cooling},
  year      = {2023},
  month     = {Sep},
  pages     = {033714},
  volume    = {108},
  doi       = {10.1103/PhysRevA.108.033714},
  issue     = {3},
  numpages  = {26},
  publisher = {American Physical Society},
  url       = {https://link.aps.org/doi/10.1103/PhysRevA.108.033714},
}

@Article{Gittes1998,
  author    = {Gittes, Frederick and Schmidt, Christoph F.},
  journal   = {Optics Letters},
  title     = {Interference model for back-focal-plane displacement detection in optical tweezers},
  year      = {1998},
  issn      = {1539-4794},
  month     = jan,
  number    = {1},
  pages     = {7},
  volume    = {23},
  doi       = {10.1364/ol.23.000007},
  publisher = {Optica Publishing Group},
}

@article{Pralle1999,
  title={Three-dimensional high-resolution particle tracking for optical tweezers by forward scattered light},
  author={Pralle, A and Prummer, M and Florin, E-L and Stelzer, EHK and H{\"o}rber, JKH},
  journal={Microscopy research and technique},
  volume={44},
  number={5},
  pages={378--386},
  year={1999},
  publisher={Wiley Online Library},
  doi       = {10.1002/(SICI)1097-0029(19990301)44:5<378::AID-JEMT10>3.0.CO;2-Z},
}

\clearpage
\newpage

\onecolumngrid

\renewcommand{\tablename}{Supplementary Table}
\renewcommand{\figurename}{Supplementary Figure}

\renewcommand{\thefigure}{S\arabic{figure}}
\renewcommand{\theequation}{S\arabic{equation}}

\setcounter{figure}{0} 
\setcounter{equation}{0}

\section*{Supplemental Material}
\pagestyle{empty}
\subsection{Distributions of $P_{\rm in}^{^{_\Omega}}(\theta_{\rm in},\phi_{\rm in})$ and $P_{\rm sc}^{^{_\Omega}}(\theta_{\rm sc},\phi_{\rm sc})$ in far-field}
\label{sec:Psc}
In the main text, we introduced a generic formula in Eq.~\eqref{eq:DeltaK2} to compute $\kappa$ based on given probability density functions. We further pointed out that these probabilities can be evaluated using the differential radiation patterns ($P_{\rm in}^{^{_\Omega}}(\theta_{\rm in},\phi_{\rm in})$, $P_{\rm sc}^{^{_\Omega}}(\theta_{\rm sc},\phi_{\rm sc})$). In this section, we provide a systematic formulation for computing these patterns.

In the most generic experimental settings the collection of tunable parameters comprises the choices of the materials of trapped objects with refractive index~($n$), the \emph{size} of the object, settings of the trapping beam such as its \emph{waist}~(${\cal O}$) and \emph{polarization}~($\vec{p}_{\rm in}$), and the \emph{focal distance}~(f) of the focusing objective. 
In this regard, evaluating $P_{\rm in}^{^{_\Omega}}(\theta_{\rm in},\phi_{\rm in})$ and $P_{\rm sc}^{^{_\Omega}}(\theta_{\rm sc},\phi_{\rm sc})$ by adjusting listed parameters becomes a tempting goal that can extend our understanding on the behavior of scattered fields upon varying these parameters.

As pointed out in the main text, we limit our analysis to the case of spherical-shaped dielectric nano-objects with complex refractive index $n$ and radius $R$. This choice is motivated by (1) the well-developed scattering theory, \emph{Mie scattering theory} (given in terms of~$n$ and~$R$ ), which provides analytical expressions employed in our calculations, and (2) the most frequent use of spherical particles in experiments on optical levitation. 
Nevertheless, we note that the employed calculation strategy has no evident limitation when it is used for exploring various object shapes (nano-dumbells~\cite{Ahn2018,Bang2020}, double-pyramids~\cite{Gosling2024}, rods~\cite{Kuhn2017} and rods aggregates~\cite{Salakhutdinov2020}). To achieve this, semi-analytical equations, as to those presented in Eq.~\eqref{eq:EscatterdMie}, should be obtained. 
 
The pattern $P_{\rm sc}^{^{_\Omega}}(\theta_{\rm sc}, \phi_{\rm sc})$ is effectively the squared amplitude of the scattered field~($E_{\rm sc}(\theta_{\rm sc},\phi_{\rm sc})$). In Mie theory, the solution for the field $E_{\rm sc}(\theta_{\rm sc},\phi_{\rm sc})$ scattered by the dielectric sphere is obtained by assuming a linearly polarized incident plane wave. However, this solution does not reflect practical conditions, as optical trapping requires a strongly focused beam, which goes beyond the paraxial regime. Thus, Mie theory has to be extended to account for the geometry of the spherical wavefront of the incident field. There are two approaches to address this point: (1) one is based on the so-called ``Generalized Mie Theory'', which requires decomposition of the incident field by vector spherical waves (see, e.g.,~\cite{Lerme2008, Gouesbet2015, RanhaNeves2019, Lamberg2023}), (2) the second one, which we employ in this work, is to decompose incident spherical wavefronts using the angular spectra of the plane waves~\cite{Toeroek1998}. The latter one stems from the work presented in~\cite{Richards1959} where the spherical wavefront was decomposed into the angular spectra of plane waves to evaluate the Debye integral (see, e.g., Sec.~8.8.1 in~\cite{BornWolf1999}) and find the vectorial electric field distribution in the vicinity of the focal region (see also discussions in~\cite{Krylenko2011}, Sec.~3.5 in~\cite{NovotnyHecht2012}).
Based on the arguments given in~\cite{Richards1959}, T{\"o}r{\"o}k et al.~\cite{Toeroek1998} proposed to extend the Mie theory over the case of an incident spherical wavefront by finding coherent superposition of the all individual Mie solutions $\sum_{k_{p}}\vec{E}_{\rm sc,k_{p}}(\theta_{\rm sc},\phi_{\rm sc})$, where each solution $\vec{E}_{\rm sc,k_{p}}(\theta_{\rm sc},\phi_{\rm sc})$ is obtained for an individual $k_{p}$-th plane wave with wavevector $\vec{k}_{ k_{p}}$ belong to the angular spectra decomposed the incident wavefront. The approach was subsequently implemented in~\cite{Rohrbach2002, Li2005, Volpe2007, Bekshaev2013}, and we refer readers to these references for the computational details as we merely elucidate the steps of the general procedure.

Let us concisely present the steps for obtaining $\vec{E}^{\rm tot}_{\rm sc}$ when a wave with spherical wavefront incident on a sphere. In Mie theory, for the case of the plane incident wave polarized along the positive direction of the x-axis, i.e., $\vec{E}_{\rm in}=E_{0, \rm in}\vec{e}_{\rm x}$, the resultant components of the electric field $\vec{E}_{\rm sc}(\theta_{\rm sc},\phi_{\rm sc})$ can be written as (see, e.g., Eq.1.6.37 in~\cite{Tsang2000})
\begin{align}
E_{\rm x, sc} (\theta_{\rm sc},\phi_{\rm sc})=& 
S_1(a_N, b_N, N)\,\sin^2\phi
\nonumber \\
&
+S_2(a_N, b_N, N)\,\cos\theta\,\cos^2\phi\,,\nonumber \\
E_{\rm y, sc} (\theta_{\rm sc},\phi_{\rm sc})=&-S_1(a_N, b_N, N)\,\sin\phi\,\cos\phi \label{eq:EscatterdMie} \\
+S_2&(a_N, b_N, N)\,\cos\theta\,\cos\phi\,\sin\phi\,, \nonumber 
\\
E_{\rm z, sc} (\theta_{\rm sc},\phi_{\rm sc})=& 
-S_2(a_N, b_N, N)\,\cos\phi\,\sin\theta\,\nonumber,
\end{align}
where $S_{i}$ with $i=1,2$ can be found in~\cite{Kerker1983}, Sec.~6.1 in~\cite{Tsang2000}, Sec.~9.3 in~\cite{Hulst1981}, and Ch.4 in~\cite{Bohren1998}. The expressions for $S_{i}$'s are presented by a sum with $N_{_{\rm max}}$ summands where each summand is a linear combinations of the \emph{scattering coefficients} $a_N({\rm R}, \lambda, n, \mu)$ and $b_N({\rm R}, \lambda, n, \mu)$ explicitly given in the same references and characterizing contributions by electric and magnetic spherical harmonics. Here, $\lambda$ denotes the wavelength of the incident light, $\mu$ sets the magnetic susceptibility of the sphere, and we assume it to be $\mu=1$. In our work, we set $N_{_{\rm max}}=5$; this value is chosen empirically, providing a relative error with respect to $N_{_{\rm max}}=20$ below $0.0005\%$. For radii exceeding 250~nm, as to in Fig.~\ref{fig:ExamplesSec5}, and for several exception points in Fig.~3 and Fig.~4 (see the main text) providing better convergence of integration in Eq.~(8) (see the main text), we set $N_{_{\rm max}}=9$. To verify the correctness of the computation of the coefficients $a_{\rm n}$ and $b_{\rm n}$, we used the reference numbers presented in Table 4.1 of~\cite{Bohren1998}, and those obtained in \cite{Maetzler2002}.

\begin{figure*}[t!]
    \centering    \includegraphics[width=0.5\linewidth]{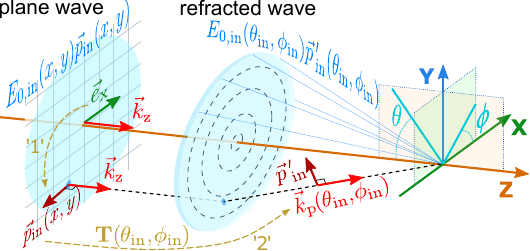}
    \caption{\textbf{Illustration of the plane-wave decomposition approach and refraction of an incident plane wave.} The transformation '1' generalizing Mie's solution in Eq.~\eqref{eq:EscatterdMie} for $\vec{e}_{\rm x}$ polarized incident plane plane wave to any arbitrary polarized $\vec{p}_{\rm in}(x,y)$ plane wave. The transformation '2' rotates polarization $\vec{p}_{\rm in}(x,y)$ to $\vec{p}\,'_{\rm in}(\theta_{\rm in}, \phi_{\rm in})$ during conversion form plane to spherical wavefront.}
    \label{fig:SchemeOfFocusing}
\end{figure*}

In the next step, we use the solution in Eq.~\eqref{eq:EscatterdMie} to find the contribution made to the total scattering $\vec{E}^{\rm tot}_{\rm sc}$ by any individual plane wave from the plane wave decomposition spectra of the incident light. Assumes the incident field is defined as $\vec{E}_{\rm in}(x,y)=E_{0, \rm in}(x,y)\vec{p}_{\rm in}(x,y)$ at the objective plane, and as  $\vec{E}_{\rm in}(\theta_{\rm in}, \phi_{\rm in})=E_{0, \rm in}(\theta_{\rm in}, \phi_{\rm in})\vec{p}\,'_{\rm in}(\theta_{\rm in}, \phi_{\rm in})$ when being refracted, see Fig.~\ref{fig:SchemeOfFocusing}. First, we establish two steps rotational transformation required to make the vector $\vec{e}_{\rm x}$ collinear to the polarization of refracted ray $\vec{p}\,'_{\rm in}(\theta_{\rm in}, \phi_{\rm in})$ (see Fig.~\ref{fig:SchemeOfFocusing})
\begin{align}
{\bm T}(\theta_{\rm in}, \phi_{\rm in}):\,\,\,  (\vec{e}_x , \vec{k}_{z}) & \xrightarrow{(\text{'1'})}  (\vec{p}_{\rm in}\,(x,y), k_{z})
\nonumber \\
&
\xrightarrow{(\text{'2'})} (\vec{p}\,'_{\rm in}(\theta_{\rm in}, \phi_{\rm in}),\vec{k}_{q} (\theta_{\rm in}, \phi_{\rm in}))\,,
\label{eq:Transformations}
\end{align}
where ($\theta_{\rm in}$, $\phi_{\rm in}$) are angular coordinates of the wavevector $\vec{k}_{q} (\theta_{\rm in}, \phi_{\rm in})$.
Second, the transformation matrix ${\bm T}(\theta_{\rm in}, \phi_{\rm in})$ introduces the rotation needed to apply to $\vec{E}_{\rm sc}$ to match it spatially with any wavevector $\vec{k}_{\rm q}(\theta_{\rm in},\phi_{\rm in})$. Thus, corresponding transformed field can be found as $\vec{E}_{\rm sc, q} = {\bm T}(\theta_{\rm in}, \phi_{\rm in})\cdot \vec{E}_{\rm sc}\big(\theta'(\theta,\phi,\theta_{\rm in}, \phi_{\rm in}),\phi'(\theta,\phi,\theta_{\rm in}, \phi_{\rm in})\big)$, where $(\theta',\phi')$ are given in a spherical coordinate in the basis where $(x',y',z')={\bm T}^{-1}(\theta_{\rm in}, \phi_{\rm in})\cdot(x,y,z)^{\intercal}$. 

To obtain $\vec{E}^{\rm tot}_{\rm sc}$, we integrate over the part of the solid angle covering the region of the incident wavefront and get
\begin{align}
    \vec{E}^{\rm tot}_{\rm sc}(\theta_{\rm sc},\phi_{\rm sc})=&A\int_{\Omega_{\rm in}}\vec{E}_{\rm sc}(\theta_{\rm sc},\phi_{\rm sc},\theta_{\rm in}, \phi_{\rm in})\, \label{eq:EscTotal}
    \\
    &
    \times E_{0,\rm in}(\theta_{\rm in}, \phi_{\rm in})\,\dd{\Omega_{\rm in}}\,\nonumber.
\label{eq:EscTotal}
\end{align}
Here, $A$ is the complex constant, which is canceled when dealing with normalized probability functions ${\mathbb P}_{\rm sc} (\theta,\phi)$ where ${\mathbb P}_{\rm sc} (\theta,\phi)=|\vec{E}^{\rm tot}_{\rm sc}(\theta_{\rm sc},\phi_{\rm sc})|^2/\int_{\Omega_{\rm sc}} |\vec{E}^{\rm tot}_{\rm sc}(\theta_{\rm sc},\phi_{\rm sc})|^2\dd{\Omega_{\rm sc}}$, with $\Omega_{\rm sc}$ covering the whole region where the light scatters into.

\begin{figure*}[t!]
    \centering
    \includegraphics[width=1\linewidth]{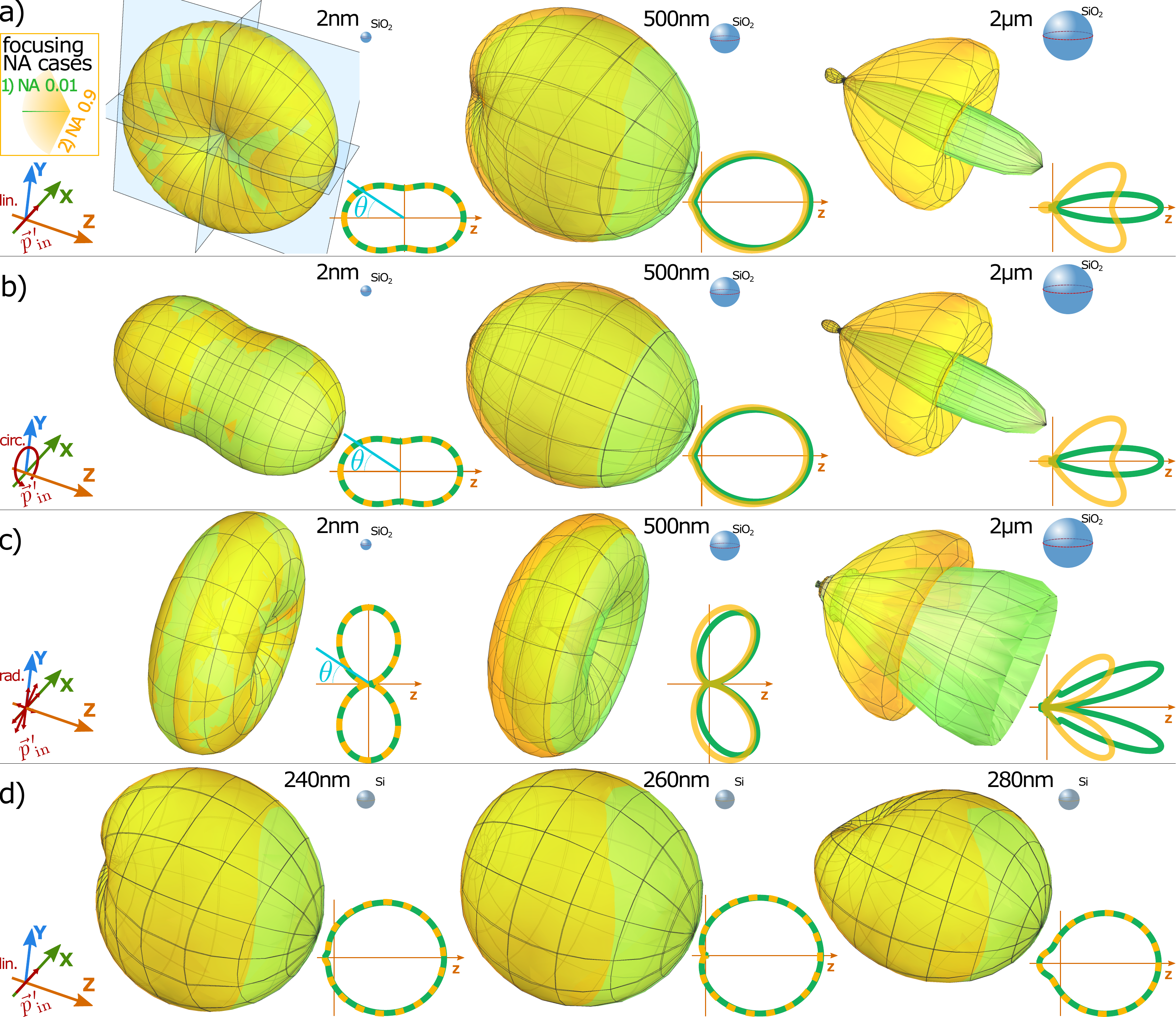}
    \caption{\textbf{Radiation patterns of the scattered light obtained for ${\rm SiO_2}$ and ${\rm Si}$ nanospheres with different diameters.} The patterns obtained as $|\vec{E}^{\rm tot}_{\rm sc}(\theta,\phi)|^{2}$ (Eq.~\eqref{eq:EscTotal}) for (a,d) linear, (b) circular, and (c) radial polarizations of the incident beam with wavelength $\lambda=1064~{\rm nm}$, filling factor $f_{0}=2$, and propagating along z-axis. The rows (a)-(c) correspond to the silica material, and row (d) shows the radiation patterns close to the conditions that lead to the Kerker effect for a silicon sphere (see Sec.~\ref{sec:KerkerEffect}). The green and orange hues correspond to the scattering of the incident beam with ${\rm NA}=0.01$ and ${\rm NA}=0.9$, respectively. The 2D plots on the lower-right corner of each 3D distribution are the azimuthally (in the x-y plane) averaged intensity distribution, and the dashed lines are used for highly overlapped curves.}
    \label{fig:ExamplesSec5}
\end{figure*}

The final stage in calculating Eq.~\eqref{eq:EscTotal} is defining the incident electric field $\vec{E}_{\rm in}$. We utilized distributions defined in the far-field for the amplitude and polarization state of the incident field. For the subsequent study presented in this work, we chose the following incident fields:
(1) Gaussian beam (see, e.g.,~\cite{Krylenko2011})
    \begin{equation}
        E^{\rm lin}_{0, \rm in}(\theta_{\rm in},\phi_{\rm in})=\sqrt{\cos\theta_{\rm in}} {\rm Exp}\left(-\frac{\sin^2\theta_{\rm in}}{f_0^2\,\sin^2 \theta_{\rm max}}\right)\,,
        \label{eq:EinGaus}
    \end{equation}
    where $f_0$ is \emph{filling factor} $f_0={\cal O}/(f\,\sin\theta_{\rm max})$, chosen to be $f_0=0.8$, and $\theta_{\rm max}=\arcsin({\rm NA})$\,. The polarization $\vec{p}_{\rm in}$ for linearly and  circularly polarized light can be written as $\vec{p}_{\rm in}=(1,0,0)^T$\, and $\vec{p}_{\rm in}=(1,i,0)^T$\,, correspondingly.
(2)) Sum of two Hermite–Gaussian modes ${\rm HG}_{\rm 01,10}$ forming radially polarized beam (see, e.g.,~\cite{Krylenko2011})
    \begin{equation}
         E^{\rm rad}_{0, \rm in}(\theta_{\rm in},\phi_{\rm in})=\sqrt{\cos\theta_{\rm in}}
        \frac{\sin{\theta_{\rm in}}}{f_0\,\sin{\theta_{\rm max}}}\,{\rm Exp}\left(-\frac{\sin^2\theta_{\rm in}}{f_0^2\,\sin^2 \theta_{\rm max}}\right)\,,
        \label{eq:EinRadial}
    \end{equation}
with $\vec{p}_{\rm in}=(\cos\phi_{\rm in},\sin\phi_{\rm in},0)^T$\, and assumed filling factor $f_0=0.7$. The factor $\sqrt{\cos\theta_{\rm in}}$ in Eq.~\eqref{eq:EinGaus} and Eq.~\eqref{eq:EinRadial} is the \emph{apodization factor} needed for energy conservation when the plane wavefront of the field transformed to the spherical wavefront under focusing with an aplanatic lens~(see, e.g.,~\cite{Davidson2004}).

\begin{figure*}
    \centering
    \includegraphics[width=0.98\linewidth]{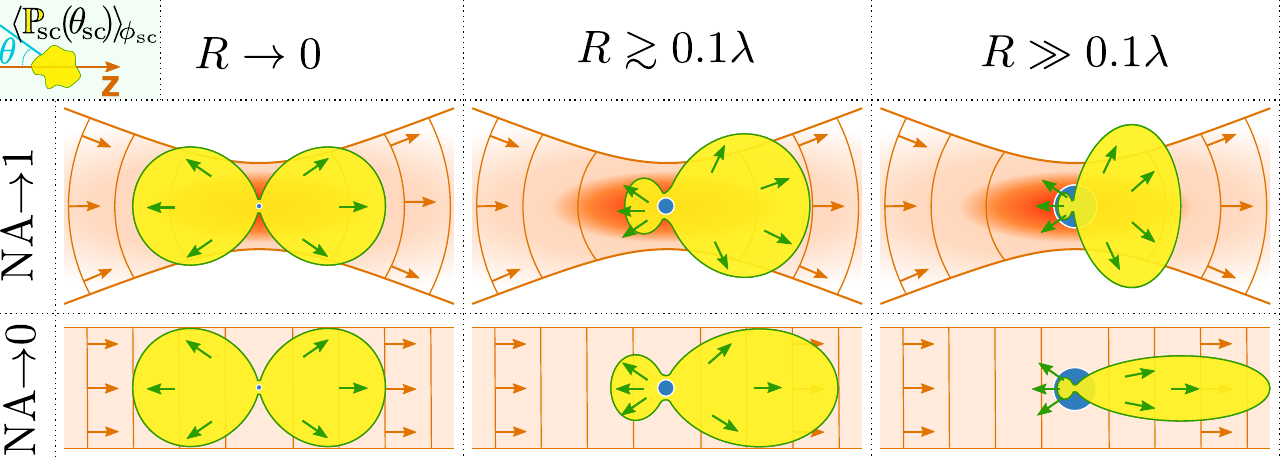}
    \caption{\textbf{Schematically drawn evolution of the radiation patterns corresponded to scatterer of different radii~($R$) and illuminated by weak and strong focused Gaussian beams}. The variation of radius is presented in columns and focusing numerical apertures in rows. The scheme is given for the radii well detuned from the Kerker condition (see SM~\cite{SuppMat}).}
    \label{fig:ConceptScheme}
\end{figure*}

The equations Eqs.~\eqref{eq:EscTotal}-\eqref{eq:EinRadial} of the incident and scattered fields serve for defining the corresponding probability density functions ${\mathbb P}_{\rm in} (\theta_{\rm in},\phi_{\rm in})$ and ${\mathbb P}_{\rm sc} (\theta_{\rm sc},\phi_{\rm sc})$. Without loss of generality, the density functions can be written as ${\mathbb P}_{\rm i} (\theta_{\rm i},\phi_{\rm i})=|\vec{E}_{\rm i}(\theta_{\rm i},\phi_{\rm i})|^2/\int_{\Omega_{\rm in}} |\vec{E}_{\rm i}(\theta_{\rm i},\phi_{\rm i})|^2\dd{\Omega_{\rm i}}$. One has to emphasize that while the scattered light assumes integration over the full solid angle $4\pi$ with $\theta_{\rm sc}\in[0,\pi]$, the integration over the angle of the incident beam is limited with the span of focusing {\rm NA} as to $\theta_{\rm in}\in[0,\arcsin({\rm NA})]$.

With the described routine to find the scattered field~(Eq.~\eqref{eq:EscTotal}), we evaluated the radiation patterns obtained for the silica~(SiO2) nanospheres of different sizes illuminated by the incident beam with $f_{0}=2$ and various polarization states, see Fig.~\ref{fig:ExamplesSec5}(a)-(c). For these figures, we choose the larger filling factor to highlight the contributions of the plane waves incident under larger angles. The selection of the silica material is driven by its wide use in optical levitation. For our calculations, we assume that the wavelength of the incident light is 1064~{nm}, and we chose the refractive index of silica to be $n_{\rm SiO_2 @ 1064nm}\approx 1.45$. First, we were interested in the role of the focusing {\rm NA} and its effect on the scattered radiation pattern. Therefore, we evaluated distributions for two contrasted {\rm NA} values, $0.01$ and $0.90$ presented in Fig.~\ref{fig:ExamplesSec5} with green and orange hues, respectively. Second, we chose three types of polarizations of the incident light, which were discussed when presenting Eqs.~\eqref{eq:EinGaus},\eqref{eq:EinRadial}, namely \emph{linear}, \emph{circular} and \emph{radial} polarizations. These cases are given in Fig.~\ref{fig:ExamplesSec5} in the panels (a), (b), and (c), respectively. Third, for the wavelength of the incident light of 1064~{nm}, we chose three different diameters of the spheres, 2nm, 500nm, and 2000nm, which are displayed in the left, middle, and right columns in panels (a), (b) and (c), correspondingly. Besides silica dielectric material, we also evaluated several patterns for silicon ${\rm Si}$ nanospheres, with complex refractive index $n_{\rm Si @ 1064nm}\approx 3.5548+82\cdot 10^{-6}\,i$, under the incidence of the linearly polarized Gaussian beam. For the three diameters (240nm, 260nm, and 280nm) in the vicinity of the size ($\approx 260$nm) where the \emph{Kerker condition} is satisfied (see Sec.~\ref{sec:KerkerEffect}), we obtained radiation patterns presented in Fig.~\ref{fig:ExamplesSec5}(d).

Based on these results, we can conclude that the sphere’s radius growth leads to the drastic radiation power redistribution between backward and forward directions. For the plane incident wave, the forward distribution forms an unidirectional angular lobe whose angular span decreases with further radius increase, gaining unidirectionality. Here, we confine our discussion with the general behavior, but the exceptional conditions available for $R\approx \lambda/5$ will be shined further in Sec.~\ref{sec:KerkerEffect}. For another variable parameter -- focusing NA, we also observe its impact becoming noticeable for a larger radius. Although, even for a relatively small silica sphere with a radius of 250~{nm}, one can notice the tendency to suppress the forward scattered radiation ($\theta_{\rm sc}=0$) and redistribute the energy to the lateral directions x, y. The high NA leads to the transformation of the unidirectional lobe to the flattened shape. To summarize the observations, we give a schematic representation of the observed behavior in Fig.~\ref{fig:ConceptScheme}, to make the above qualitative observations more comprehensible.

\subsection{Kerker effect}
\label{sec:KerkerEffect}
In this section, we discuss the \emph{Kerker effect} (see~\cite{Kerker1983,Liu2018}), which causes the light scattering to be predominantly forward-directed, and examine the asymmetry factor that characterizes this directivity. The idea behind the Kerker effect (see, e.g.,~\cite{Nechayev2019}) lies in control and optimization of the magnitude and phase of the complex coefficient $a_{\rm N}$, $b_{\rm N}$, discussed in Sec.~\ref{sec:Psc}, in order to suppress the backward~(forward) scattering light. The straightforward condition to realize the Kerker effect and suppress backward scattering (so-called \emph{first Kerker condition}) is attaining the equality of complex magnitudes of the electric and magnetic dipoles, namely $a_{_{\rm N=1}}=b_{_{\rm N=1}}$. This condition leads to the destructive interference between these two contributions along the negative valued $z$ axis and constructive interference along the positive valued $z$ axis, with the object placed at $z=0$.

To demonstrate the role of the interference between the electric field generated by magnetic and electric sources induced in a sphere, we use the \emph{asymmetry parameter} $\langle \cos \theta \rangle$ (see, e.g., Sec.~3.11~\cite{Kerker1969}, Sec.~2.3 in~\cite{Hulst1981}, Sec.~4.5~\cite{Bohren1998}). The asymmetry parameter is the mean magnitude of $\cos \theta_{\rm sc}$ weighted by the weighting function $\mathbb{P}_{\rm sc}(\theta_{\rm sc},\phi_{\rm sc})$ such that 
\begin{equation}
\langle \cos \theta_{\rm sc} \rangle_{_{\rm \Omega}} =\int_{\rm \Omega_{\rm sc}}\mathbb{P}_{\rm sc}(\theta_{\rm sc},\phi_{\rm sc}) \cos\theta_{\rm sc}\,\dd \Omega_{\rm sc}\,, 
\label{eq:CosT}
\end{equation}
and represent the balance between back- and forward-scattered light. The cases $\langle \cos \theta_{\rm sc} \rangle<0$ and $\langle \cos \theta_{\rm sc} \rangle>0$ correspond to light scattered more in the direction opposite or along the z-axis, respectively. As it follows from Eq.~\eqref{eq:CosT}, the magnitude of $\langle \cos \theta_{\rm sc} \rangle$ as high as directivity of the scattered light and get its maximum for the case $\mathbb{P}_{\rm sc}(\theta_{\rm sc},\phi_{\rm sc})=\delta(\theta_{\rm sc}-\pi)$ (in the coordinate system we chose, see Fig.~\ref{fig:SchemeOfFocusing}). For the spherical-shaped dielectric scatterer and linear polarized incident plane wave, the asymmetry parameter can be calculated as follows:

\begin{equation}
    \langle \cos \theta_{\rm sc} \rangle_{_{\rm \Omega}}=\frac{2\,\displaystyle \sum _{{\rm N}=1}^{N_{_{\rm max}}}\left( \frac{1+2{\rm N}}{{\rm N}(1+{\rm N})}{\rm Re}[a_{\rm N}\,b_{\rm N}^{*}]+\frac{{\rm N}(2+{\rm N})}{(1+{\rm N})}{\rm Re}[a_{\rm N}\,a_{\rm N+1}^{*}+b_{\rm N}\,b_{\rm N+1}^{*}] \right)}{ \displaystyle \sum _{{\rm N}=1}^{N_{_{\rm max}}} (2 {\rm N}+1) (|a_{\rm N}|^2+|b_{\rm N}|^2)},
    \label{eq:CosMean}
\end{equation}
where coefficients $a_{\rm N}$ and $b_{\rm N}$ were presented in Sec.~\ref{sec:Psc} being introduced in Eqs.~\eqref{eq:EscatterdMie}, and we set $N_{_{\rm max}}=5$ as in the general computations discussed in Sec.~\ref{sec:Psc}.

Evaluation of the asymmetry factor and finding its local maxima allow us to find the conditions that provide the minimal radiation pressure, that leads to lowering recoil energy and parameter $\kappa_{\rm z}$. In this regard, we calculated the magnitude of asymmetry factor $\langle \cos \theta \rangle$ given in Eq.~\eqref{eq:CosMean} for the range of real-valued refractive index $n \in [1.2,\,3.7]$ and radii $R\in$[1~{nm},\,250~{nm}] assuming the wavelength of the incident light to be of $1064\,{\rm nm}$. The obtained distribution presented in Fig.~\ref{fig:Cos} in the main text allows us to find the radius of the scatterer for any non-absorbing materials providing significant forward scattering, and represented as highest magnitudes of the $\langle \cos\theta_{\rm sc}\rangle$. To tackle the role of the reduction of $\kappa$, in the current work, we use three materials: silica~(${\rm SiO_{2}}$), silicon (Si), and diamond (Di). We emphasize that any high refractive material can be used for such procedure in general (see, e.g.,~\cite{NietoVesperinas2010}). For the given materials with refractive indexes $n_{\rm Di @ 1064nm}\approx 2.39$ and $n_{\rm Si @ 1064nm}\approx 3.5548+82\cdot 10^{-6}\,{\rm i}$, we have the maxima of asymmetry factor at $R_{\rm Di}\approx 193\,{\rm nm}$ and $R_{\rm Si}\approx 130\,{\rm nm}$. To demonstrate the important role of the radius choice, we plot the radiation pattern of the Si nanosphere near the radius of the maximum forward scattering $R_{\rm Si}=130\pm10~{\rm nm}$. From Fig.~\ref{fig:ExamplesSec5}(d), it is evident that even a slight change in the particle size affects the symmetry of the radiation pattern.

\subsection{Scattering cross-section}
\label{sec:cross-section}

\begin{figure}[t!]
    \centering
    \includegraphics[width=0.5\linewidth]{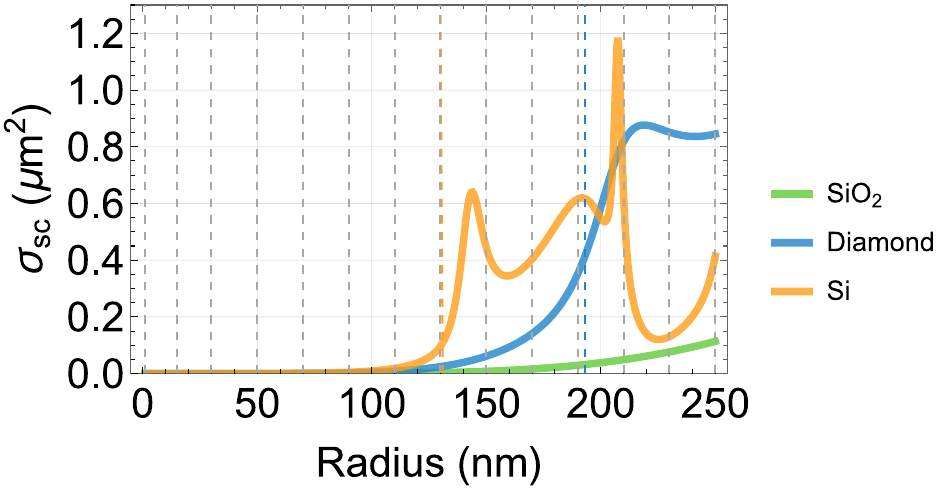}
    \caption{\textbf{Scattering cross-section $\sigma_{sc}$ calculated by Eq.~\eqref{eq:SigmaPlane} for the plane linearly polarized incident wave.}
    The distributions plotted for silica~(green), diamond~(blue), and silicon~(orange) nanospheres as a function of the particles' radii. The gray dashed lines correspond to the equidistant radii magnitudes chosen to plot Fig.~\ref{fig:SigmaSCNA}. The blue and dashed lines correspond to the radii of the maximal asymmetry factor obtained by Eq.~\eqref{eq:CosMean} for diamond and silicon, correspondingly.
    }
    \label{fig:SigmaSCplane}
\end{figure}

The cross-section ($\sigma_{\rm sc}$) of Mie particles illuminated by a linearly polarized plane wave is given by equation (see, e.g., Sec.~4.4.1~\cite{Bohren1998} and Sec.~6.1 in~\cite{Tsang2000})
\begin{equation}
\sigma_{\rm sc}=\frac{2\pi}{k^{2}}\sum _{{\rm N}=1}^{N_{_{\rm max}}}(2{\rm N}+1)\left(|a_{\rm N}|^{2}+|b_{\rm N}|^{2}\right)\,,
\label{eq:SigmaPlane}
\end{equation}
coefficients $a_{\rm N}$ and $b_{\rm N}$ are the same as in previous sections. Figure \ref{fig:SigmaSCplane} displays $\sigma_{\rm sc}$ as a function of radius for silica, diamond, and silicon nanospheres obtained by Eq.~\eqref{eq:SigmaPlane}. This equation can also approximate the results for cross-sections when Mie particles are illuminated by a weak focused beam, linearly polarized wave with small NA. Therefore, the results presented in Fig.~\ref{fig:SigmaSCplane} serve us as a benchmark with the reference magnitudes for low NA cases.

\begin{figure*}
    \centering
    \includegraphics[width=0.98\linewidth]{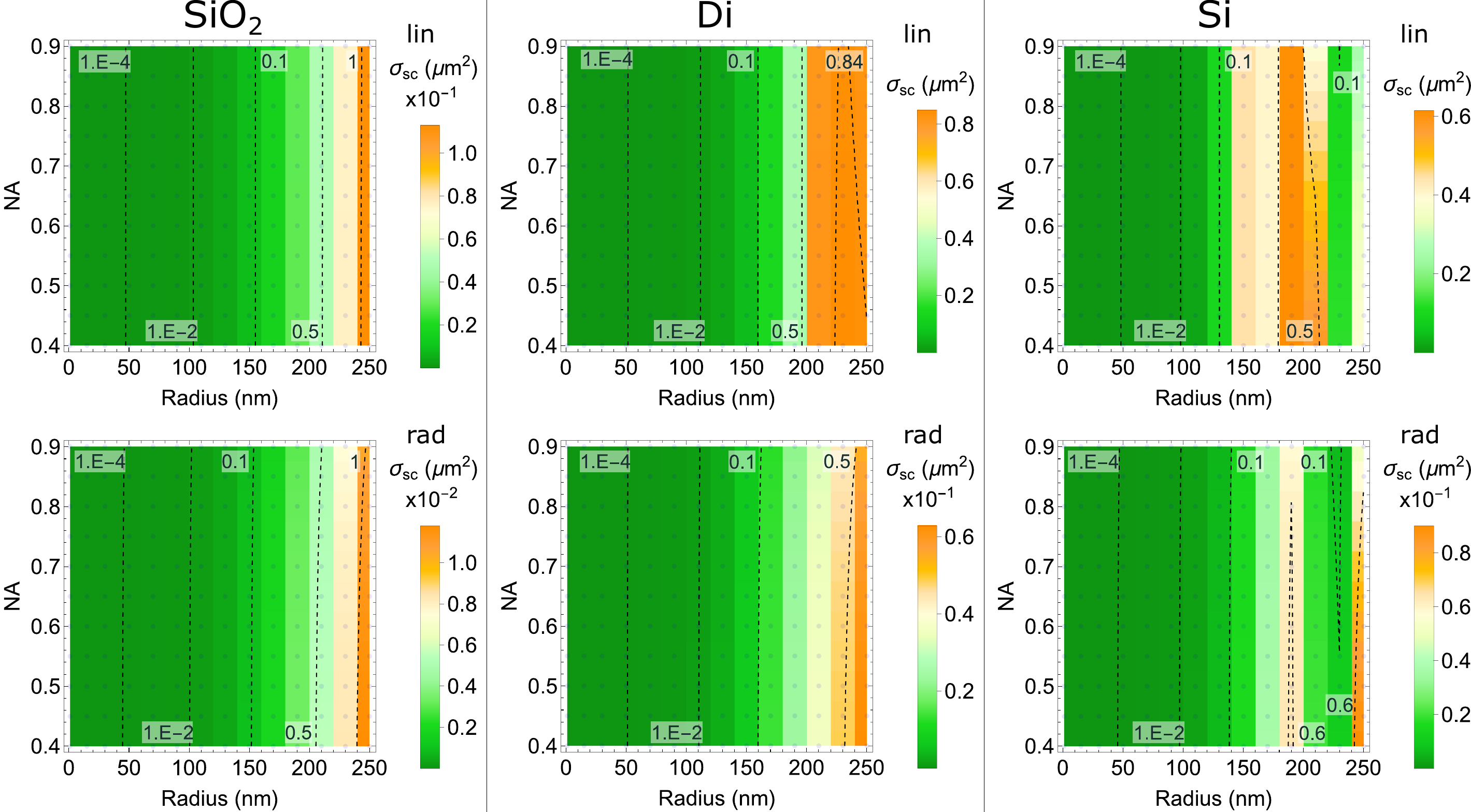}
    \caption{\textbf{Scattering cross-section $\sigma_{sc}$ calculated with Eq.~\eqref{eq:scatteringGeneral}.} Columns from left to right correspond to silica, diamond, and silicon materials, correspondingly. The top (bottom) row corresponds to the linearly (radially) polarized focused beam. The contour lines are obtained by interpolating the values calculated with Eq.~\eqref{eq:scatteringGeneral} at the coordinates marked with dark points.}
    \label{fig:SigmaSCNA}
\end{figure*}

Since Eq.~\eqref{eq:SigmaPlane} can not be used directly for the scattering processes involving high NA, and does not account for the radial polarization of the incident beam, the alternative approach to calculate the cross-section has to be used. For the low-absorbing particles considered in this work, the scattering cross-section $\sigma_{\rm sc}$ is equivalent to the extinction cross-section $\sigma_{\rm ext}$ and can be found via the optical theorem as
\begin{equation}
\sigma_{\rm sc}\approx\sigma_{\rm ext}=-\frac{4\pi}{k}{\rm Im}[\,\vec{p}_{\rm in}\cdot\vec{E}_{\rm sc}(\theta=\pi)]\,.
\label{eq:OpticalTheorem}
\end{equation}
However, this definition is only valid for incident beams with defined polarization states and the nonzero scattered light along the optical axes. Since the radially polarized modes have undefined polarization at their center, and since the scattering of radially polarized light by Mie particle results in a zero electric field along the optical axis (see,e.g., Fig.~\ref{fig:ExamplesSec5}(c)), using of the Eq.~\ref{eq:OpticalTheorem} is impossible (see, e.g.,~\cite{Krasavin2018}). To overcome this limitation, we calculated $\sigma_{\rm sc}$ using numerically calculated distributions of $|\vec{E}^{\rm tot}_{\rm sc}(\theta_{\rm sc},\phi_{\rm sc})|$ and substituting them in the general definition of the scattering cross-section (see, e.g., Sec.~1.1 in~\cite{Tsang2000}, Sec.~3.4 in~\cite{Bohren1998} and~\cite{Krasavin2018}):
\begin{equation}
    \sigma_{\rm sc} = \frac{\int_{\Omega_{\rm sc}}P_{\rm sc}(\theta_{\rm sc},\phi_{\rm sc})\dd{\Omega_{\rm sc}}}{I_{\rm in, f}}\,.
    \label{eq:scatteringGeneral}
\end{equation}
Here $P_{\rm sc}(\theta_{\rm sc},\phi_{\rm sc})=\frac{\epsilon_0 c}{2}|\vec{E}^{\rm tot}_{\rm sc}(\theta_{\rm sc},\phi_{\rm sc})|^2$, and $I_{\rm in, f}$ is intensity of the incident field at exact position of the focus $(0,\,0,\,0)$, which is defined as $I_{\rm in, f}=\frac{\epsilon_0 c}{2}|\vec{E}_{\rm in, f}|^{2}$. The electric fields $\vec{E}_{\rm in, f}$ for linearly and radially polarized focused beams were obtained by calculations presented in~\cite{Krylenko2011} with the following expressions obtained at the focal point coordinates:
\begin{align}
\vec{E}^{\rm lin}_{\rm in, f}=&\vec{E}^{\rm lin}_{\rm in, x}(0,0,0)= \int\limits_{\Omega_{\rm in}}  E^{\rm lin}_{0, \rm in}(\theta_{\rm in}) (1+\cos{\theta_{\rm in}})\sin{\theta_{\rm in}}\,\dd \theta_{\rm in}\,,\\
\vec{E}^{\rm rad}_{\rm in, f}=&\vec{E}^{\rm rad}_{\rm in, z}(0,0,0)= 2\int\limits_{\Omega_{\rm in}}  E^{\rm rad}_{0, \rm in}(\theta_{\rm in}) \sin^2{\theta_{\rm in}}\,\dd \theta_{\rm in}\,,
\end{align}
where all other components are equal to zero for the respective cases. The obtained distributions of $\sigma_{\rm sc}$ in dependence on the radius of the particle and focusing NA are presented in Fig.~\ref{fig:SigmaSCNA} and were plotted by spline function of the first order interpolating results obtained in the coordinates marked with dark points.

We compared the results in Fig.~\ref{fig:SigmaSCNA}~(top row), obtained for the incident linear polarization, with those obtained for the plane incident wave given in Fig.~\ref{fig:SigmaSCplane}. This comparison was performed across the entire set of calculated points and was consistent for the selected radii. The peaks at about 146~{nm} and 207~{nm} corresponded to silicon and are not resolved in Fig.~\ref{fig:SigmaSCNA} due to the grid we have chosen. 

The observed agreement for linearly polarized incident beams validates the approach used with Eq.~\eqref{eq:scatteringGeneral}. This consistency gives us confidence in applying the same method to obtain the scattering cross-section for radially polarized beams, as shown in Fig.~\ref{fig:SigmaSCNA}~(bottom row).

\subsection{Selected model and its applicability}
To evaluate the consistency of our approach within conventional frameworks, we first consider Eq.~(8) in the dipole-scatterer limit. Expanding the brackets, we obtain:
\begin{equation}
\kappa_{\rm i}= \langle \hat{k}^{2}_{\rm in,i}\rangle_{_{\Omega_{\rm in}}}
-2\,\langle \hat{k}_{\rm in,i}\cdot \hat{k}_{\rm sc,i}\rangle_{_{\Omega_{\rm in},\Omega_{\rm sc}}}
+\langle \hat{k}^{2}_{\rm sc,i}\rangle_{_{\Omega_{\rm sc}}}.
\end{equation}
Assuming that the two events are independent (see section Methods in the main text), the second term on the right-hand side reduces to the product 
$\langle \hat{k}_{\rm in,i}\rangle_{_{\Omega_{\rm in}}} \langle \hat{k}_{\rm sc,i}\rangle_{_{\Omega_{\rm sc}}}$.  
Furthermore, because the angular spectrum of the scattered wave considered here is symmetric with respect to positive and negative directions of $k_{\rm i}$, the expression for a dipolar scatterer simplifies to
\begin{equation}
\kappa_{\rm i}= \langle \hat{k}^{2}_{\rm in,i}\rangle_{_{\Omega_{\rm in}}}+\langle \hat{k}^{2}_{\rm sc,i}\rangle_{_{\Omega_{\rm sc}}}.
\end{equation}

\begin{figure}[t!]
    \centering
    \includegraphics[width=1\linewidth]{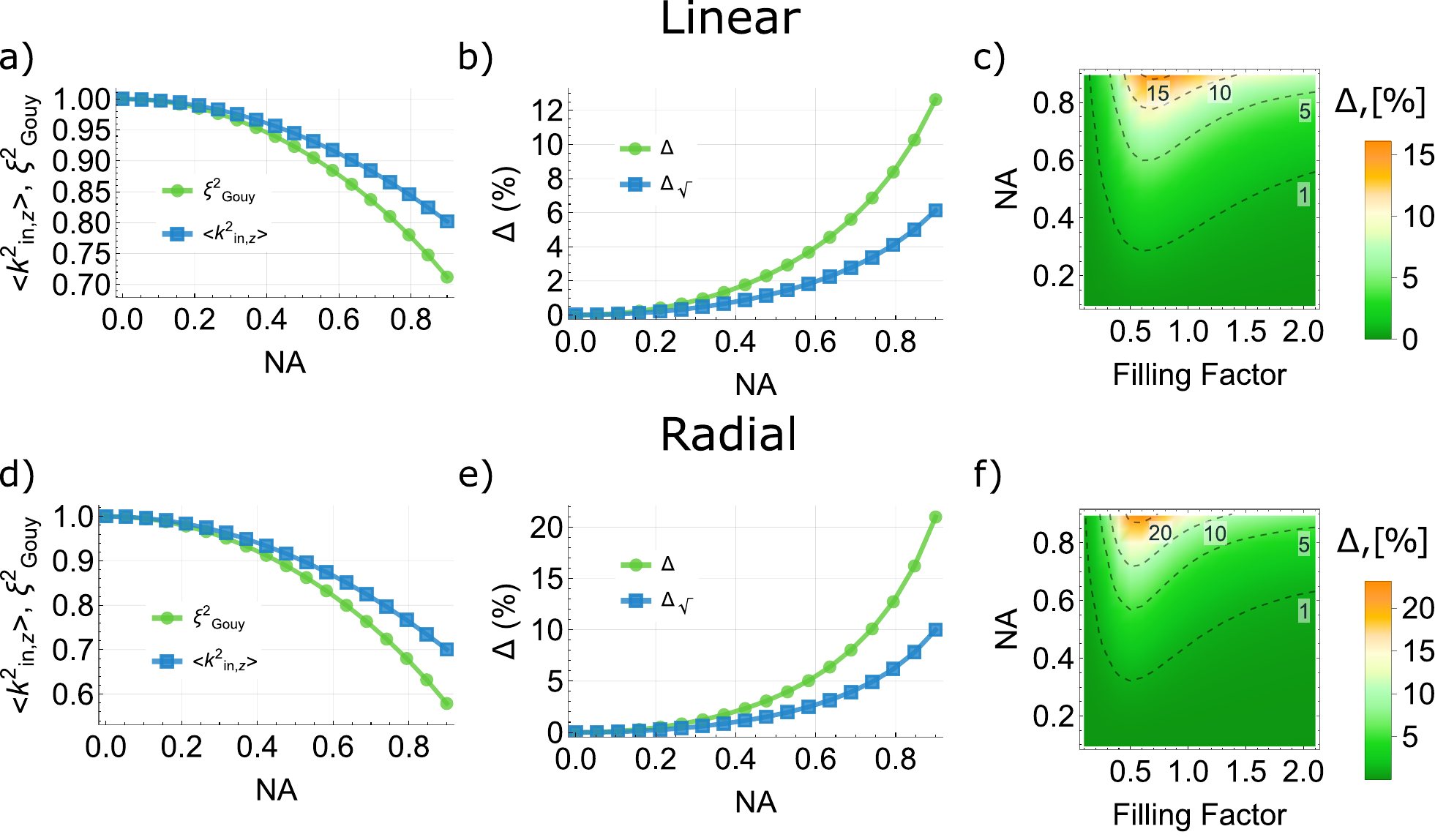}
    \caption{\textbf{Quantitative comparison between $\langle \hat{k}^{2}_{\rm in,z}\rangle_{_{\Omega_{\rm in}}}$ (far-field) and $\xi^2_{_{\rm Gouy}}$ (near-field).}
Panels a,d show $\langle \hat{k}^{2}_{\rm in,z}\rangle{\Omega_{\rm in}}$ and $\xi^2_{{\rm Gouy}}$ for linearly and radially polarized incident light with filling factors $f{0}=0.8$ and $f_{0}=0.7$, respectively. Panels b,e present the corresponding relative differences calculated according to Eqs.~\eqref{eq:DiffKz}, \eqref{eq:DiffKzSQR}, while c,f display the relative difference $\Delta$ for a broad range of filling factors.
    }
    \label{fig:ToleranceLinRad}
\end{figure}

In the far-field regime, averaging over $\Omega_{\rm in}$ is performed with respect to the intensity of the angular spectrum. Consequently, the average value $\langle k_{\rm in,z} \rangle$ obtained in this way differs from the parameter $\xi_{_{\rm Gouy}}$ defined in the near-field through amplitude averaging (see Eq.~A3 in~\cite{Tebbenjohanns2019}), where the isotropic source is replaced by the actual angular spectrum amplitude. To clarify this difference, both quantities $\langle \hat{k}^{2}_{\rm in,z}\rangle$ and $\xi^2_{_{\rm Gouy}}$ (as in Eq.~6 in the main text) were calculated for the incident beams considered in our work. For a linearly polarized Gaussian beam and a radially polarized beam focused by a lens with ${\rm NA}=0.9$ and with $f_0=0.8$ and $f_0=0.7$, respectively, the results are shown in Fig.~\ref{fig:ToleranceLinRad}~(a,~d). The relative differences between the far-field and near-field values for these cases are presented in Fig.~\ref{fig:ToleranceLinRad}~(b,~e), where they are expressed as percentages,
\begin{align}
\Delta=&\frac{\langle \hat{k}^{2}_{\rm in,z}\rangle_{\Omega_{\rm in}}-\xi^2_{_{\rm Gouy}}}{\xi^2_{_{\rm Gouy}}}\cdot 100\%\,,\label{eq:DiffKz}\\
\Delta_{\sqrt{ }}=&\frac{\sqrt{\langle \hat{k}^{2}_{\rm in,z}\rangle_{\Omega_{\rm in}}}-\sqrt{\xi^2_{_{\rm Gouy}}}}{\sqrt{\xi^2_{_{\rm Gouy}}}}\cdot 100\%\,.\label{eq:DiffKzSQR}
\end{align}
These plots demonstrate that averaging with the angular spectrum intensity, as applied in our work, overestimates the energy component along the $z$ direction during photon absorption by approximately $13\%$ and $21\%$. However, these discrepancies are expected to represent nearly the maximum deviations for the dipole approximation across the broad ranges ${\rm NA}\in[0.1,0.9]$ and $f_0\in[0.1,2.1]$.

The ability to calculate fields in the focal region (\cite{Krylenko2011} and Sec.~3.5 in~\cite{NovotnyHecht2012}) allows one to evaluate the Gouy phase gradient and thereby determine the parameter $\xi_{_{\rm Gouy}}$ across a wide range of numerical apertures and filling factors, for both linear and radial polarizations. The corresponding relative differences for $\langle k^2_{\rm in,z}\rangle$ ($\Delta$ given with Eq.~\ref{eq:DiffKz}) are shown in Fig.~\ref{fig:ToleranceLinRad}~(c,~f). Analysis of these results indicates that the maximum discrepancy between the far-field and near-field models is about $16\%$ and $23\%$ for the case of a dipole scatterer illuminated by linear and radial polarizations, respectively.

Having discussed the limitations of our approach in the dipole approximation, we now turn to its applicability in the general case. For the total superposition of the incident and scattered fields, $\vec{E}_{\rm in}(\theta,\phi)+\vec{E}_{\rm sc}(\theta,\phi)$, the balance of the net energy flux through a sphere of radius $R \gg \lambda$ in the absence of absorption can be expressed as the sum of integrals over the sphere surface for each term in $|\vec{E}_{\rm in}(\theta,\phi)+\vec{E}_{\rm sc}(\theta,\phi)|^2$ (see Sec.~13.5.3~\cite{BornWolf1999}), i.e. 
$W_{\rm in}+W_{\rm int}+W_{\rm sc}$, where ``int'' stands for the interference term $\vec{E}_{\rm in}\cdot(\vec{E}_{\rm sc})^*$. In the case of a low-absorbing medium, $W_{\rm in}\approx0$ (see Sec.~13.5.3~\cite{BornWolf1999}), and its contribution to the trapping is associated not with scattering forces but with the gradient force (see, e.g., ~\cite{Rohrbach2002,Rohrbach2004}). Thus, the resulting expression for the energy flux balance in integral form becomes
\begin{equation}
\underbrace{\int\limits_{\Omega\in4\pi} \vec{E}_{\rm in}(\theta,\phi)\cdot(\vec{E}_{\rm sc}(\theta,\phi))^*\dd\Omega}_{W_{\rm int}}+\underbrace{\int\limits_{\Omega\in4\pi} |\vec{E}_{\rm sc}(\theta,\phi)|^2\dd\Omega}_{W_{\rm sc}}\approx0
\label{eq:EnergyFluxIntegration}
\end{equation}
which corresponds to the identity Eq.~\eqref{eq:OpticalTheorem} since $\sigma_{\rm ext} = W_{\rm int}/I_{\rm in,f}$ by definition. The sum of these integrals, taken as a projection on the normal to the sphere surface in each direction, yields an expression for the average $i$-th component of the scattering force obtained via treating the Maxwell stress tensor in the far-field approximation (see, e.g.,~\cite{Novotny2017}).

The integral of $|\vec{E}_{\rm sc}|^2$ in Eq.~\eqref{eq:EnergyFluxIntegration}, evaluated using the expression (see Eq.~\eqref{eq:EscTotal}), coincides with our expression $\mathbb{P}_{\rm sc}(\theta_{\rm sc},\phi_{\rm sc})$ obtained from the angular spectrum of the scattered radiation. This term can therefore be regarded to the scattered contribution. The remaining task is to evaluate how well our approach captures the contribution of the interference term in Eq.~\eqref{eq:EnergyFluxIntegration} to the characterization of the incident light $\mathbb{P}_{\rm sc}(\theta_{\rm sc},\phi_{\rm in})$. For this purpose, we use the identity valid for a particle located exactly at the focus:
\begin{equation}
W_{\rm int}=\int\limits_{\Omega\in4\pi} \vec{E}_{\rm in}(\theta,\phi)\cdot(\vec{E}_{\rm sc}(\theta,\phi))^*\dd\Omega=\int\limits_{\Omega\in4\pi} \vec{E}_{\rm in}(\theta,\phi)\Big(\int\limits_{\Omega_{\rm in}} f_{\rm sc}(\theta_{\rm sc},\phi_{\rm sc},\theta, \phi)\,\vec{E}_{\rm in}(\theta_{\rm in}, \phi_{\rm in})\,\dd{\Omega_{\rm in}}\Big)^*\dd\Omega\,,
\end{equation}
where $f_{\rm sc}$ is the scattering amplitude which is the general representation of the transformations given in Eq.~\eqref{eq:EscTotal}. Since we consider the far-field approximation without angular coherence, i.e., $f_{\rm sc}(\theta_{\rm sc},\phi_{\rm sc},\theta, \phi)=f_{\rm sc}(\theta, \phi)\delta(\theta-\theta_{\rm sc},\phi-\phi_{\rm sc})$, the integration reduces to the form
\begin{equation}
W_{\rm int}\approx \int\limits_{\Omega\in4\pi} \vec{E}_{\rm in}(\theta,\phi) \Big( f_{\rm sc}(\theta, \phi)\,\vec{E}_{\rm in}(\theta, \phi)\Big)^*\dd\Omega=\sigma_{\rm ext} \int\limits_{\Omega\in4\pi} |\vec{E}_{\rm in}(\theta,\phi)|^2\dd\Omega\,.
\end{equation}
This expression reflects the essence of the approximation used in our approach with Eq.(1), where the focal intensity of the incident light $I_{\rm in,f}$ is approximated by the integral of the incident field power over the angular spectrum, i.e., $I_{\rm in,f}\propto\int|\vec{E}_{\rm in}(\theta, \phi)|^2\dd\Omega$. This approximation is a rough estimate compared to the near-field approach, where  $I_{\rm in,f}\propto|\int\vec{E}_{\rm in}(\theta, \phi)\dd\Omega|^2$, based on the Debye integrals given in Eq.(8), and this difference constitutes the source of the error discussed earlier in this section, leading to the discrepancy shown in Fig.~\ref{fig:ToleranceLinRad}.

The near-field model approximates the incident wavefront as flat and therefore neglects the transverse wave-vector components, an assumption valid only at subwavelength distances. In contrast, our approach treats the incident radiation as angularly incoherent, consistent with conditions away from the focal region, and naturally accounts for the redistribution of energy flux between longitudinal and transverse directions, thereby exceeding the capabilities of the near-field model. Since it relies on Mie theory with a plane-wave decomposition, this framework provides an accurate description of the scattering process for particles beyond the Rayleigh approximation, as previously demonstrated using a similar method for calculating the scattered field (see, e.g.,~\cite{Bekshaev2013}). We therefore conclude that our approximation is most reliable for particles beyond the dipolar regime, with its upper validity ultimately constrained by the applicability of Mie scattering theory.

\end{document}